\documentclass[preprint,floatfix, showpacs,superscriptaddress]{revtex4}
\usepackage{color,graphicx}
\usepackage{bm}
\usepackage{amsmath}
\usepackage{amssymb}
\usepackage{float}

\usepackage{changes}

\newcommand{\be}{\begin{eqnarray}}
\newcommand{\ee}{\end{eqnarray}}

\begin{document}
\title{Quantum-Critical Fluctuations in 2D Metals: 
Strange Metals \\ and Superconductivity in Antiferromagnets and in Cuprates} 
\author{Chandra M. Varma}
\affiliation{Department of Physics and Astronomy, University of California, Riverside CA 92521, USA}

\begin{abstract}
The anomalous transport and thermodynamic properties in the quantum-critical region, in the cuprates, and in the quasi-two dimensional Fe-based superconductors and heavy-fermion compounds, have the same temperature dependences. This can occur only if, despite their vast microscopic differences, a common statistical mechanical model describes their phase transitions. The antiferromagnetic (AFM)-ic models for the latter two, just as the loop-current model for the cuprates, map to the dissipative XY model. The solution of this model in 2+1 D reveals that the critical fluctuations are determined by topological excitations, vortices and a variety of instantons, and not by renormalized spin-wave theories of the Landau-Ginzburg-Wilson type, adapted by Moriya, Hertz and others for quantum-criticality. The absorptive part of the fluctuations is a separable function of momentum ${\bf q}$, measured from the ordering vector, and of the frequency $\omega$ and the temperature $T$ which scale as $\tanh(\omega/2T)$ at criticality. Direct measurements of the fluctuations by neutron scattering in the quasi-two-dimensional heavy fermion and Fe-based compounds, near their antiferromagnetic quantum critical point, are consistent with this form. Such fluctuations, together with the vertex coupling them to fermions, lead to a marginal fermi-liquid, with the imaginary part of the self-energy  $\propto max(\omega, T)$ for all momenta, a resistivity $\propto T$, a $T \ln T$ contribution to the specific heat, and other singular fermi-liquid properties common to these diverse compounds, as well as to d-wave superconductivity. This is explicitly verified, in the cuprates, by analysis of the pairing and the normal self-energy directly extracted from the recent high resolution angle resolved photoemission measurements. This reveals, in agreement with the theory, that the frequency dependence of the attractive irreducible particle-particle vertex in the d-wave channel is the same as the irreducible particle-hole vertex in the full symmetry of the lattice. 
 \end{abstract}
\date\today
\maketitle
\section{Introduction and Phenomenology} 

\begin{figure}[tbh]
\centering
\includegraphics[width=1.0\columnwidth]{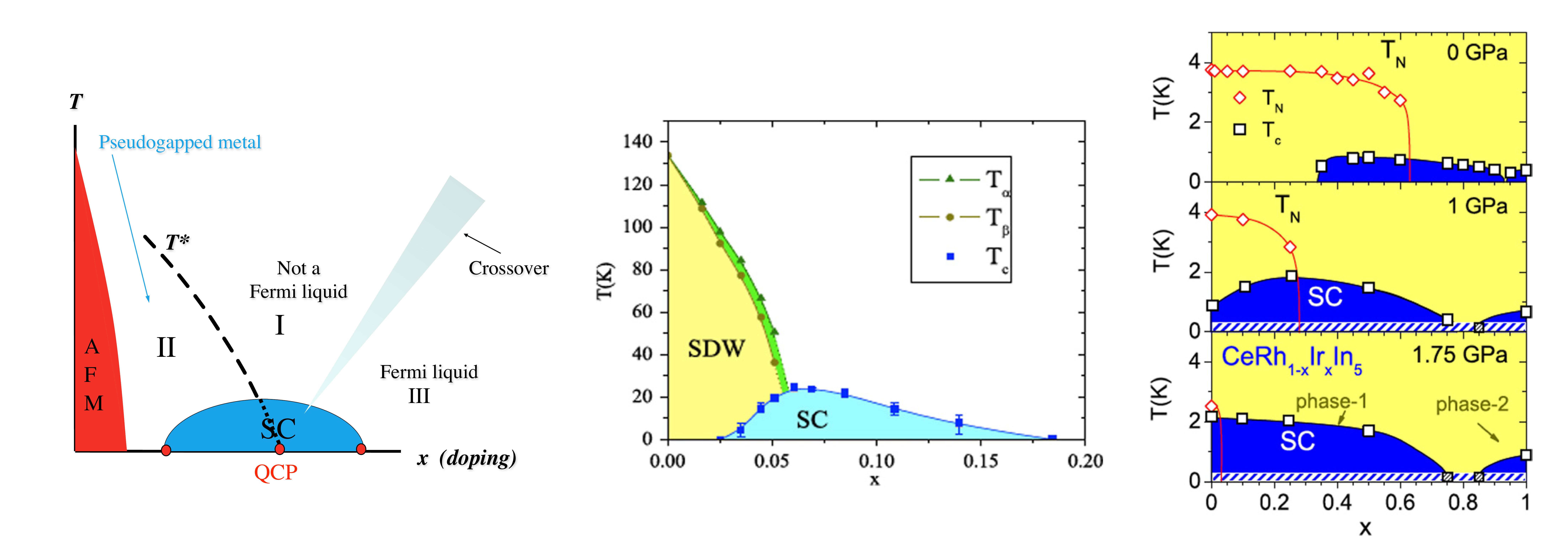}
\caption{Left: The universal Phase Diagram for hole-doped cuprates. The spin-glass region and the weak charge density waves with and without magnetic fields, which occur in Region II are not shown.
Nor is the termination of the $T^*(x)$ line.  Middle: The phase diagram of Ba(Fe$_{1-x}$Co$_x$)$_2$As$_2$ \cite{BaAsFe, Canfield2009}; the green triangles mark the transition to an altered structure while the black circles mark the AFM transition. The superconducting region is shown in blue. Right: The phase diagram of the indicated heavy-Fermion compounds, taken from \cite{HFPhDia}.}
\label{Fig:CommonPhDia}
\end{figure}

\begin{figure}[tbh]
\centering
\includegraphics[width=1.0\columnwidth]{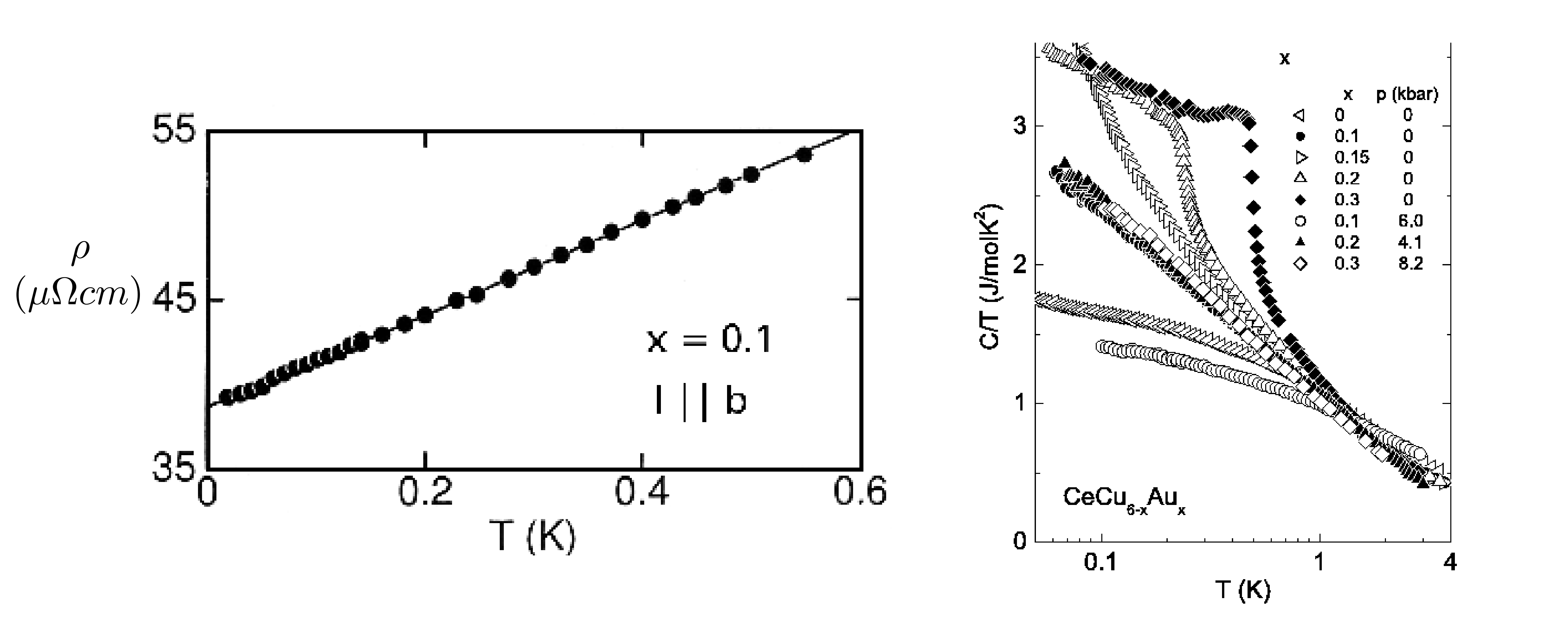}
\caption{ Left: Resistivity of CeCu$_{5.9}$Au$_{0.1}$ and Right: Specific heat of  CeCu$_{6}$ under various substitution of Au or under pressure, near AFM quantum-criticality is shown; from \cite{HvL1996}}
\label{Fig:RCCe}
\end{figure}

\begin{figure}[tbh]
\centering
\includegraphics[width=1.0\columnwidth]{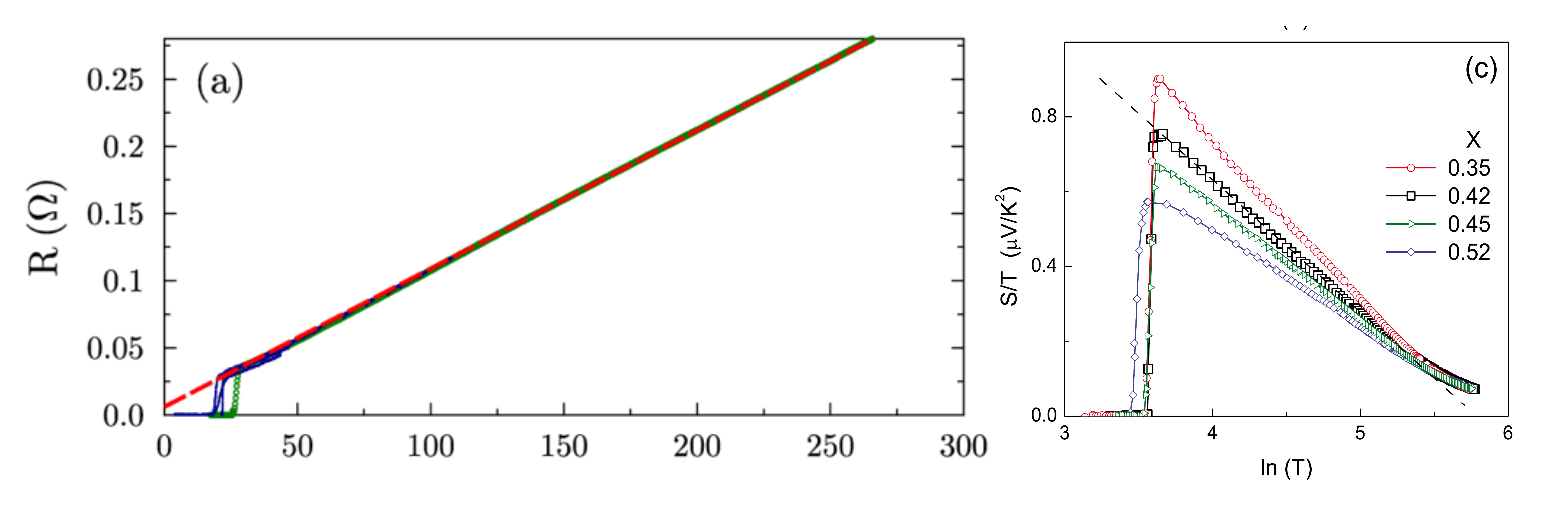}
\caption{ Left: Resistance at various magnetic fields from 0 to 65 Tesla, given in different colors, in Ba(Fe$_{1-x}$Co$_x$)$_2$As$_2$ for $x=0.41$ close to the quantum-critical composition, from \cite{Analytis1}, Right: Thermopower in K$_x$Sr$_{1- x}$Fe$_2$As$_2$ from \cite{ThemopowerFe}}
\label{Fig:RS}
\end{figure}

The schematic phase diagram of a Fe-based superconductor, a heavy-fermion compound and of the hole-doped cuprates is shown in Fig. (\ref{Fig:CommonPhDia}).
All three are anisotropic 2 D metals; the first two, in the region of critical fluctuations due to their (AFM)-ic quantum-critical point (QCP),  have properties remarkably similar to those in cuprates in a region I of their phase diagram. 
The dependence of the resistivity on temperature in both is linear, the entropy or thermopower (proportional to entropy per carrier) is proportional to $T \ln T$, and the nuclear relaxation rate, where available, has a large constant part. The temperature dependence of the resistivity and the specific heat in CeCu$_6$ for various substitutions of Au for Cu or under pressure, near quantum-criticality \cite{HvLRMP2007, CMV_Lorentz}, is shown in Fig. (\ref{Fig:RCCe}). The resistance in one of the Fe-based compounds near quantum-criticality, with and without magnetic fields, and the thermopower are shown in Fig. (\ref{Fig:RS}). The Fe compounds in the Fermi-liquid region of their phase diagram have an effective mass of O(1), while the well studied heavy Fermion compound, CeCu$_{6-x}$Au$_x$ \cite{HvLRMP2007} or CeCu$_{6-x}$Ag$_x$ \cite{GegenwartPRL2004} have an effective mass of $O(10^3)$.  All these anomalies (and many more) were discovered \cite{Ginzburg-rev} and studied \cite{Anderson-book, CMV-MFL} first in the cuprates in region I of Fig. \ref{Fig:CommonPhDia}, which is obviously not a region of AFM criticality but abuts the region II, which occurs along a line $T^*(x)$, where thermodynamic and transport properties change universally in all cuprate families.

A magneto-electric order due to loops of orbital currents, breaking time-reversal and reflection symmetry but preserving translational symmetry, was predicted for region II of 
Fig. (\ref{Fig:CommonPhDia}) in cuprates \cite{cmv1997, simon-cmv}.  There is considerable direct experimental support for such a symmetry \cite{Bourges-rev, Greven, Kaminski-diARPES}. No other phase has been discovered starting at $T^*(x)$ in the extensive experimental investigations in the cuprates. The statistical mechanical model \cite{Gronsleth-PRB2009} for the loop-current order is the Ashkin-Teller model which has critical fluctuations on the disordered side akin to those of the 2D XY model. The classical phase transition in the XY-model in 2D is of the Kosterlitz-Thouless \cite{KT}, Berezinsky \cite{Berezinskii} kind, which is determined by statistical mechanics of topological excitations - the vortices, and not by renormalized spin-waves as in the usual Ginzburg-Landau-Wilson (GLW) type of phase transitions. The quantum-critical fluctuations in such a model, supplemented with dissipation of the collective degrees of freedom \cite{Aji-V-qcf1, Aji-V-qcf3, ZhuChenCMV2015} are also determined by topological excitations, vortices, and instantons of phase jumps in time, termed 'warps'. 
 
 Unlike the cuprates, where the order parameter in Region II is novel, there is no question but that  the quantum-critical fluctuations in the heavy-fermions and in the Fe-based compounds are the fluctuations of AFM order. (A tetragonal to orthorhombic transition often accompanies the AFM transition in the Fe compounds, as in Fig. (\ref{Fig:CommonPhDia}). But there is no evidence of the quantum fluctuations of such a transition in scattering the fermions.) The similarity of the normal state anomalies in the diverse systems in Fig. \ref{Fig:CommonPhDia} suggested an inquiry as to whether the model for AFM quantum criticality maps also to that of the XY model. This depends on symmetry. If the fluctuations are isotropic in spin-space or if the correlation length perpendicular to the planes is not negligible compared to those in the plane, no such mapping is possible. But in real materials, due to crystal fields and spin-orbit coupling, the effective magnetic moments are often anisotropic and so are the exchange energies in spin-space. The eventual magnetic order is either uni-axial or along one of the several equivalent directions in a plane. 
If they are also spatially very anisotropic, the fluctuation regime has a region dominated by 2D uni-axial or planar magnetic fluctuations.
In Sec. (II), we briefly review the result that under these conditions, incommensurate uni-axial fluctuations, and both commensurate or incommensurate planar magnetic fluctuations, map to the (2+1)D XY model. 
The topological excitations, vortices and warps, map to topological defects of the antiferromagnets - vortices to edge-dislocations and warps to flip in time of local sub-lattice magnetization. 

A common theoretical framework, the quantum dissipative 2D XY model, will therefore be discussed for the quantum-critical properties of the cuprates and the anisotropic heavy fermion compounds, and the Fe-based anti-ferromagnets/superconductors.

The dynamical version \cite{Hoh-Hal-RMP} of the Ginzburg-Landau-Wilson theory of classical phase transitions was adopted for quantum phase transitions by Moriya \cite{Moriya-book}, Hertz \cite{Hertz}, Beal-Monod and Maki \cite{Beal-Monod} and by others. In this theory, the characteristic frequency of the fluctuations, $\omega$ scales as ${\bf q}^z$, where ${\bf q}$ is the deviation of the wave-vector of the fluctuations from the ordering vector ${\bf Q}$, and the $z$, the dynamical critical exponent depends on the symmetry of the order parameter and the nature of the dissipation. Such a theory, and its many variations, has not yielded results consistent with the observed anomalies in the quantum-critical region of any of the classes of compounds we have mentioned. The idea \cite{Woelfle2011} that vertex corrections in the Moriya-Hertz theory may change the  singularities  in transport do not lead to the linear in T resistivity. The idea \cite{SiNature2001} that heavy-fermion quantum-criticality is related to the breakdown of a Kondo impurity \cite{Larkin_Melnikov, Maebashi1, Maebashi2} in a self-consistent bath  needs justification that the theory is stable against the interactions of Kondo 'impurities'. It has also not given the linear in T resistivity. 

The theory of the phase transitions of the dissipative quantum XY model in 2D \cite{Aji-V-qcf1,  ZhuChenCMV2015} is in a different class from the GLW type theories. The correlations are separable in space and time, with the spatial correlation length proportional to the logarithm of the temporal correlation length. The concept of a dynamical critical exponent $z$ is not useful. At criticality, the absorptive part of the fluctuations in frequency $\omega$ and temperature $T$ are proportional to $\tanh{(\omega/2T)}$. Correspondingly the real part is $\propto \log (max(\omega, T))$. There is not a "soft mode" but a continuum of critical modes from the lowest frequency to the ultra-violet cut-off. This frequency and temperature dependence was suggested long ago in a phenomenological model \cite{CMV-MFL} to explain the anomalous properties of the cuprates, but the derived momentum dependence was not guessed earlier. In the microscopic theory, the vertex coupling the fermions to the fluctuations has also been calculated. Using the momentum dependence of the vertices,  the anomalies in thermodynamics and transport in Region I in Fig. \ref{Fig:CommonPhDia}, follow also from the derived form of the fluctuations.  The form of the vertices also turns out to be essential to understand the symmetry of the superconductivity, and of the frequency dependence of the normal self-energy and of the pairing self-energy in the superconducting state. These have been deduced directly from angle-resolved photo-emission (ARPES) experiments recently \cite{Bok_ScienceADV}. One may expect to encounter a similar situation when such experiments are available in Fe-based superconductors.  

In using these results to interpret experiments, it must be remembered that the coupling in direction perpendicular to the plane is a relevant perturbation to the classical 2D XY model. Although, this is likely to be the case for the quantum model as well, an important point in this connection is that, as we will present, the in-plane spatial correlation length in the dissipative (2+1)D XY model only increases logarithmically as the critical point is approached. One should expect therefore, that the marginal fermi-liquid quantum-critical properties (resistivity linear in T, etc.) may hold up to very close to the critical point unless three-dimensional couplings are strong and other perturbation such as the effect of impurities are relevant.

The mapping of the models for phase transitions in the AFM's and the cuprates to the XY model is summarized in Sec. II.  The coupling of the fluctuations to the fermions, necessary to discuss dissipation in the XY model, the fermion self-energy, and the effective pairing interactions are reviewed in Sec. III.  Aspects of the derivation of the fluctuations of the dissipative XY model are presented in Sec. (IV).  An analysis of the fluctuation spectra directly measured by inelastic neutron scattering experiments gives results which appear to verify the unusual predictions. These are reviewed in Sec. (V), as well as the singular self-energy of fermions in the normal state by scattering the fluctuations.
How such fluctuations lead to the observed superconductivity is reviewed in Sec. (VI), where  evidence for such fluctuations in cuprates will be summarized as well as a discussion of the interesting new problems posed by the Fe-based compounds. 

The same model is applicable to quantum superconductor-insulator transitions \cite{Goldman-SIT-rev, ChakraKivel1}, to ferromagnetic quantum critical transitions \cite{AronsonPNAS2014} and by some accounts, the plateau transitions in quantum Hall effects \cite{kivel-lee-zhang}. In this review we do not discuss these problems.

\section{Mapping to the XY Model}

{\it Uni-axial AFM}: In this case, the mapping to the XY model follows the arguments for incommensurate charge density waves \cite{OverhauserCDW, McMillanCDW}. Suppose the anisotropies in a metal are such that the magnetic order and the fluctuations in the quantum-critical region are at an incommensurate vector ${\bf Q}$ and have the symmetry of uni-axial spin fluctuations. The order parameter is 
$M_z e^{i({\bf Q \cdot R}_i + \phi)}$,
where $M_z$ is the amplitude, ${\bf Q}$ is the incommensurate wave-vector, and ${\bf R}_i$ are lattice sites. Any uniform value of $\phi$ is then allowed since this amounts only to a change of the zero of the co-ordinates. The leading cost in Free-energy due to variations of the order parameter between neighboring sites is 
\be
J \sum_{(i,j)} |M_{z,i}e^{i({\bf Q \cdot R}_i + \phi_i)} - M_{z,j}e^{i({\bf Q \cdot R}_j + \phi_j)} |^2.
\ee
Assuming, as shown below, that the model for of $\phi$ is an XY model, the variations of the amplitude $M$ are unimportant. Then, an effective Hamiltonian  for small variations of $\phi$ is
\be
J a^2 M_z^2 \int d{\bf r}~ \lambda_{\|} |(\nabla_{\|} \phi- i{\bf Q})|^2 +  \lambda_{\bot}|\nabla_{\bot} \phi|^2, 
\ee
where 
$\nabla_{\|, \bot}$ are derivatives parallel and perpendicular to ${\bf Q}$ and the $\lambda$'s account for the differences in stiffness parallel and perpendicular to ${\bf Q}$. This form ensures that the minimum in spin-density variation is at wave-vector ${\bf Q}$. For larger variations, the potential energy can only depend periodically on the difference of phase $(\phi_i-\phi_j)$.  Therefore, the uniaxial incommensurate AFM fluctuations are described by an XY model for the phase $\phi$. The correlation functions in the AFM at $({\bf q-Q})$  are the same as the the correlation functions of the ferromagnetic XY model (or superfluid) at ${\bf q}$. In 2D, the classical critical fluctuations are determined by the correlation functions of vortices. The edge dislocations in the incommensurate AFM in 2D correspond to vortices in 2D superfluids. 

 {\it Planar AFM}:
In this case, the model for fluctuations around a commensurate wave-vector on a bi-partite lattice maps to the XY model (with lattice anisotropy). This is seen by  gauge transformation on a planar-AFM Hamiltonian, whereby the spin on alternate sites is reversed as well as the sign of the coupling. The model is then that for a FM-XY model. 

The order parameter for the incommensurate AFM is
\be
{\bf M}(\theta) e^{({\bf Q \cdot R}_i + \phi)},
\ee
${\bf M}(\theta)$ is now a vector in the plane, at an angle $\theta$ with respect to, say ${\bf Q}$. Now both uniform $\phi$ and $\theta$ can have any value. So the model would appear to be in the $U(1) \times U(1)$ class. But consider the cost in energy of their variations between sites.  Again, neglecting variations in the amplitude of ${\bf M}$, the effective Hamiltonian for small variations in $\phi$ and $\theta$ is
\be
J a^2 M^2 \sum_{(i,j)} \lambda_{\|} |(\nabla_{\|} - i{\bf Q})\phi + \nabla_{\|}{\bf M}(\theta)|^2 +  \lambda_{\bot}|\nabla_{\bot} \big(\phi + {\bf M}(\theta)\big)|^2, 
\ee
where only the gradients of the angular variations of ${\bf M}(\theta)$ are to be taken. The gradients of such variations are linearly coupled to the gradients of $\phi$. Therefore it is impossible to have independent fluctuations or topological defects 
of $\phi$ and of $\theta$. In particular, dislocations in the $\phi$ field must be accompanied by vortices in the $\theta$ field. This is similar to a coupled case,  treated by Nelson \cite{Nelson}. The transition is of the Kosterlitz-Thouless variety with a change in the exponents from the case of only one periodic variable.

\begin{figure}[tbh]
\centering
\includegraphics[width=0.8\columnwidth]{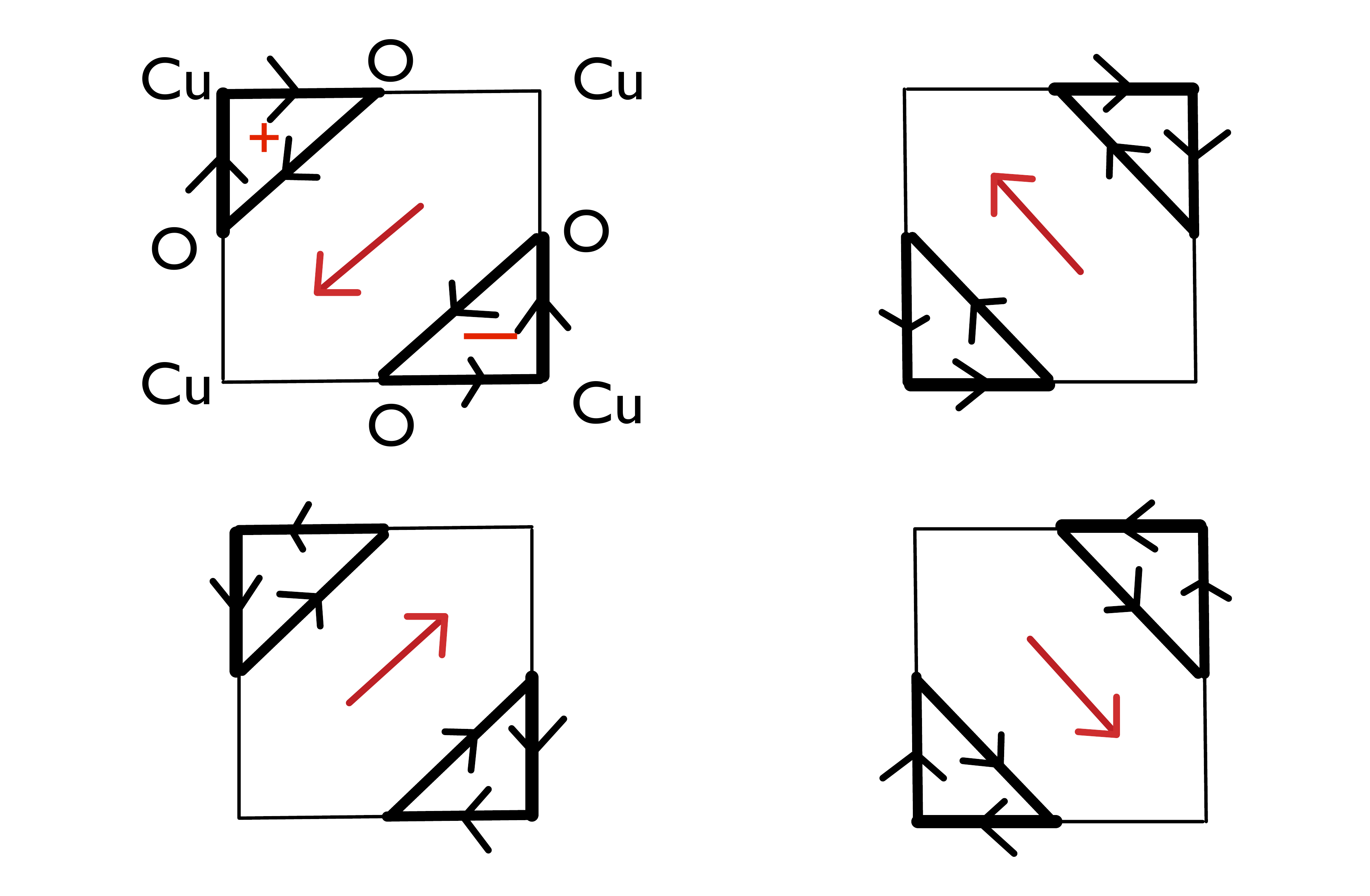}
\caption{The four possible orientations of the time-reversal and inversion breaking order parameter ${\bf \Omega}$, suggested for the cuprates for Region II of Fig \ref{Fig:CommonPhDia}. There are two closed -current loops per unit-cell with a resulting magnetic moment along $\hat{z}$ and another along $-\hat{z}$, in any of the four possibilities, so that ${\bf \Omega}$'s  are the vectors in red. The quantum-critical region is characterized by fluctuations of ${\bf \Omega}$ among the four-orientations, so that quantum XY model, with four-fold anisotropy is used to describe the fluctuations}
\label{Fig:Cuprate-OP}
\end{figure}

 {\it Loop-Current Order in Cuprates}: 
 Unlike the case for the heavy-fermions and the Fe-based AFM/superconductors, the phase for cuprates in region II of Fig. \ref{Fig:CommonPhDia} has not been obvious, nor easy to discover, even though the change in the transport and thermodynamic and transport properties due to it suggest that a substantial fraction of the fermionic degrees of freedom are affected by it. It was proposed that the transition to region II, at $T^*(x)$, occurs due to ordered loops of currents within a unit-cell, without breaking translational symmetry. This phase can be characterized by time-reversal and reflection symmetry breaking (magneto-electric) phase characterized by the anapole vector, 
\be
\label{Om}
{\bf \Omega}_i = \int_{cell-i} d^2r~ ({\bf M}_i({\bf r}) \times {\bf r}). 
\ee 
${\bf M}_i({\bf r})$ is the magnetization density at ${\bf r}$ in the unit-cell $i$. ${\bf \Omega}_i$ has four orientations in the plane, as exhibited in Fig. \ref{Fig:Cuprate-OP}. 
 Translational symmetry of moments consistent with an ordered phase of loop-currents has been found below the pseudogap temperature
$T^*(x)$ in four different families of cuprates, for which there are large enough single crystals to do polarized neutron diffraction measurements \cite{Bourges-rev, Greven}.  There are some general questions about polarized neutron scattering raised in these experiments by the disagreement in the direction of moments from their interpretation as classical moments perpendicular to the planes. This issue has been addressed by constructing a theory of polarized neutron diffraction for quantum moments \cite{He-V-neutrons}. Dichroic ARPES experiments, which rely on time-reversal and the specific reflection symmetries broken by such a phase were proposed \cite{cmv-dichroicARPES}. In the compound Bi2212, the signatures of such a phase have been observed \cite{Kaminski-diARPES} consistently below the $T^*(x)$ measured by other experiments.  An unusual birefringence in light propagation occurs due to the magneto-electric order, in which the principal axes for propagation of polarized light themselves rotate as the order parameter increases.\cite{Armitage-Biref, CMVEurophys2014}. Time-reversal breaking leading to an unusual Kerr effect occurs \cite{Kapitulnik1, Kapitulnik2, cmv-kerr} also as observed, in such a phase, if some reflection symmetries are also independently broken. The magneto-electric phase characterized by ${\bf \Omega}$ has the further virtue that this is the only phase identified to occur starting at $T^*(x)$, determined from the thermodynamic and transport properties, in any of the families of hole-doped cuprates in any of the many and varied experiments carried out.

There is no divergence in specific heat in the Ashkin-Teller or XY model at the phase transition; there is a very large fluctuation regime and a weak non-analytic feature in the specific heat at the transition \cite{Gronsleth}. The singular feature is too weak to be discerned directly \cite{CMV-Zhu-PNAS} in specific heat measurements. The variation with temperature of the sound velocity, which is more accurately measurable, and is proportional to the specific heat \cite{CMV-Zhu-PNAS}, do see both the large fluctuation regime and the weak non-analytic feature \cite{Shekhter2013} consistent with the expectations. Features in the magnetic susceptibility, consistent with loop-current order have also been consistently found at $T^*(x)$ \cite{LeridonSUSC}.

Local slow probes, such as $\mu$SR, and NMR have however not seen such a phase. It was proposed \cite{CMV-Domains2014} that this may be due to domains of this phase with quantum fluctuations at a rate faster than $10^{-5}$ secs. characteristic  of such experiments. This issue is still unsettled. 

The magnitude of the measured order \cite{Bourges-rev}, about  $0.1 \mu_B/$unit-cell, counting both current loops, at O$_{6.87}$ to $0.2 \mu_B$/unit-cell at O$_{6.6}$ in YBa$_2$Cu$_3$O$_{y}$ and the temperature of its occurence can be used to calculate the reduction in energy \cite{CMV-Zhu-PNAS} due to such orders. For an ordered moment of $0.1 \mu_B/$unit-cell, the energy reduction is similar to the superconducting condensation energy, about 50 Joules/mole,
 at the largest transition temperature as a function of doping. Therefore the loop-current state is a candidate as a competing state for superconductivity and for providing sufficient amplitude of quantum-critical fluctuations. From this point of view the weak magnitude charge density wave states seen in several cuprates are non-starters.

\section{The XY Model with Interaction with Fermions}

\begin{figure}[tbh]
\centering
\includegraphics[width=1.0\columnwidth]{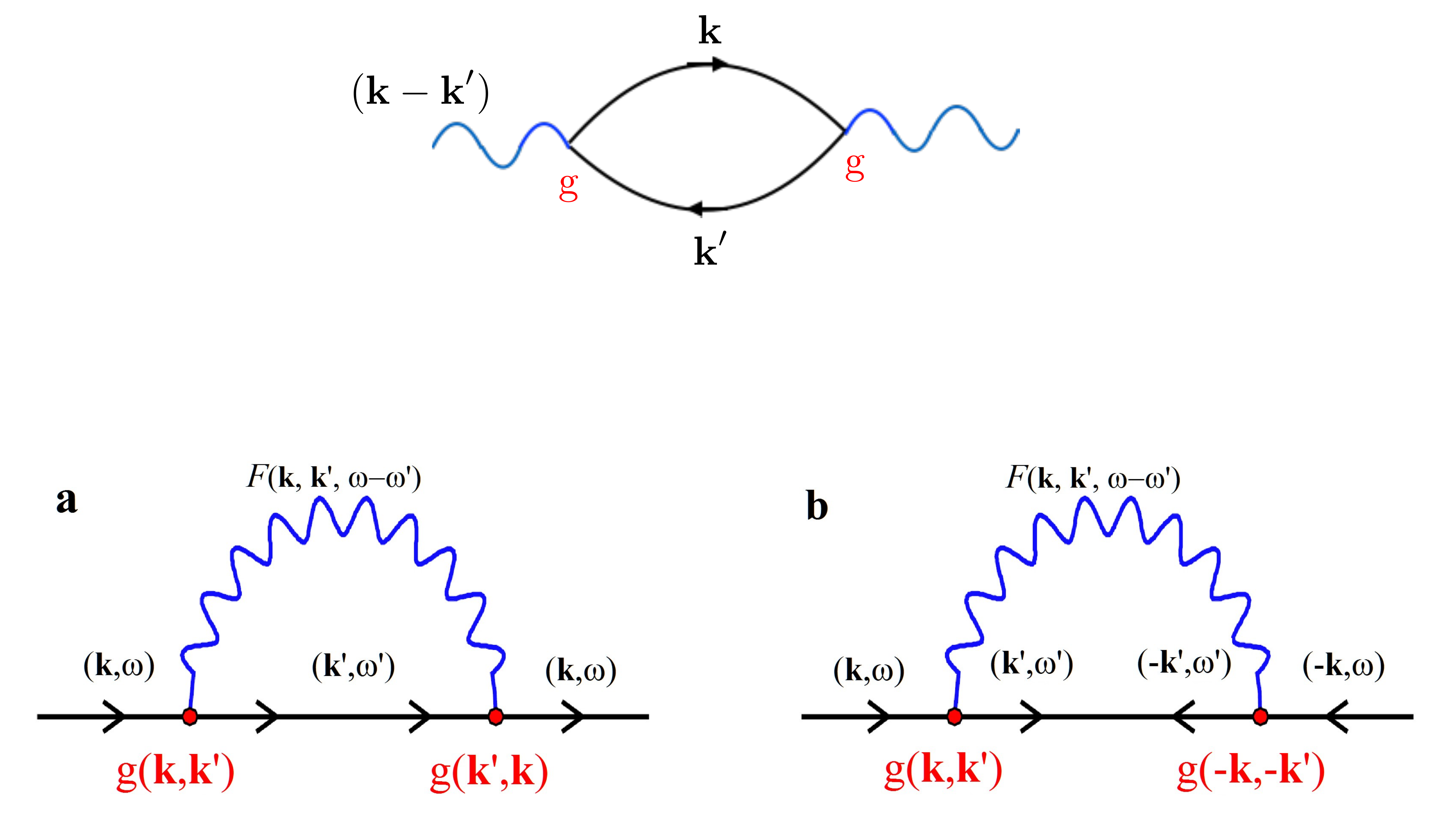}
\caption{ Top: Skeleton diagram with vertices coming from the coupling of the collective fluctuations to the fermions, providing a dissipation for the former (wiggly line in blue) by dissipating into fermion-currents due to finite zero temperature resistivity. Bottom: (a) Skeleton diagram for the normal self-energy and (b) for the pairing self-energy with the same vertices.}
\label{Fig-damp-self-energy}
\end{figure}

The strategy of solution adopted in this class of problems, AFM-ic or loop-currents, is to start with a fermion Hamiltonian with appropriate interactions, identify the important collective degrees of freedom, and deduce a Hamiltonian through Hubbard-Stratonovich or equivalent transformations, which has the form,
\be
H =H_F + H_C + H_{CF}.
\ee
$H_F$ is a non-interacting Fermion Hamiltonian, and $H_C$ is the Hamiltonian for the collective degrees of freedom, which in the present instance map to the XY model. $H_C$ consists of the potential energy of two-dimensional rotors, $H_{pot}$ and their kinetic energy $H_K$.  $H_{CF}$ is the interaction Hamiltonian for the  Fermions to scatter off the collective fluctuations of the rotors. $H_{CF}$ serves two purposes - it provides dissipation to the collective degrees of freedom through processes shown as the skeleton diagram in Fig (\ref{Fig-damp-self-energy}-top) and renormalizes the Fermions through process shown in Fig (\ref{Fig-damp-self-energy}-bottom), providing both the normal (a) and the pairing self-energy (b).

 In 2D, the fluctuations of the magnitude of the rotors are not relevant and one may take fixed length rotors interacting with each other through the potential energy:
\be
H_{pot} = \sum_{i(j)} -K \cos(\theta_i - \theta_j) + \sum_i h_4 \cos(4\theta_i)
\ee
The second term describes a four-fold lattice anisotropy. For the classical model, anisotropy is irrelevant if more than 4-fold and marginal at 4-fold \cite{KT1973}. It is shown in perturbative calculations  \cite{Aji-V-qcf3} that the anisotropy is irrelevant in the quantum problem. Monte-carlo calculations \cite{ZhuChenCMV2015} give the same results in the critical fluctuation regime with and without the anisotropy term, for $h_4/K_0$ up to at least 4. 
Therefore, we will drop the anisotropy term hereafter.
 For the proposed broken symmetry in the cuprates, $\theta_i$'s are just the angles of the anapoles $\Omega_i$ at the cell $i$. For the 2D antiferromagnet, we will consider the equivalent model for 2D-superfluidity, so that $\theta_i$'s are the superfluid phases at a lattice point $i$. 

The kinetic energy of the rotors is,
\be
\label{KE}
H_K = \sum_i \frac{1}{2C} L^2_{z,i},
\ee
where $L_{z,i} = i \partial/\partial \theta_i$ is the angular momentum, conjugate to $\theta_i$ and $C$ is their moment of inertia.

Near the phase transitions in a metal, part of the spectral weight of the electronic excitations is converted to that of the critical fluctuations while the rest remains as incoherent excitation. Then it is essential also to consider dissipation from conversion of the former to the latter. In order to do so, we must first derive the coupling of the collective degrees of freedom, the $\theta_i$'s to fermions.

{\it Coupling to Fermions}

The minimum coupling of the Fermions to the collective fluctuations of the quantum XY model to the fermions comes in two varieties:\\
(i) Coupling of the phase fluctuations to the fermions: \\ 
Only the gradient of the phase,which is proportional to the current due to the collective fluctuations can couple, and it can couple only to the current operator of the fermions:
\be
\label{coupling1}
H_{CF}^{(1)} = \int d{\bf r} g_0~ \nabla \theta ({\bf r})\cdot\psi^+_{\sigma}({\bf r}){\bf J}\psi_{\sigma}({\bf r}) +H.C. \\ \nonumber
= \sum_{{\bf k,q}, \sigma}i g_0~ \theta({\bf q})  \frac{{\bf q}\cdot (2{\bf k + q})}{m}\psi^+_{{\bf k+q}, \sigma}\psi_{{\bf k}, \sigma}
\ee
It can be shown that (\ref{coupling1}) is transformed to a coupling between AFM fluctuations and fermions \cite{CMV-PRL2015}. This is precisely of the form of dissipation which is introduced in the LGW-type theory of AFM- quantum critical fluctuations \cite{Hertz}.

One may also wish to keep the coupling $e^{i(\theta({\bf r}) - \theta({\bf r}'))} \psi^+_{\sigma}({\bf r})\psi_{\sigma}({\bf r}')$ so that periodicity is maintained in ${(\theta({\bf r}) - \theta({\bf r}'))}$. We have not not found a procedure to do analytical calculations with dissipation introduced by such a term. It is however found in Monte-Carlo calculations that, dissipation introduced in this manner has no effect in the phase transition in the XY model, when dissipation introduced through coupling of the form (\ref{coupling1}) is also present even when the coupling constant in the former is as much as 5 times larger than the latter. The periodic coupling represents dissipation of vortices while the form (\ref{coupling1}) is due to dissipation of small spin-wave like fluctuations. \\

\noindent
(ii) Coupling through the angular momentum $L_z$: \\
The coupling between the collective modes and the incoherent fermion excitations through the angular momentum $L_z$ of the rotors of (\ref{KE}) is also important. This coupling has been derived microscopically in the case of the cuprates \cite{ASV2010, He-V-CollModes}. It can also be written on general symmetry grounds. Local angular momentum of the fluctuations can couple only to the local angular momentum operator of the fermions. So, the coupling, in the continuum approximation, is necessarily of the form
\be
\label{coupling2}
H_{CF}^{(2)} = \int d{\bf r} \sum_{\sigma}g_0'  ~{\bf L}_z({\bf r}) \psi^+({\bf r}, \sigma)\frac{1}{2}({\bf r}\times {\bf p} - {\bf p} \times {\bf r}) \psi({\bf r}, \sigma). \\ \nonumber
= \sum_{{\bf k,q}, \sigma} i g_0' ~ F(|{\bf k}|)~{\bf L}_z({\bf q}) \cdot   ({\bf k} \times{\bf q}) \psi^+_{{\bf k+q}, \sigma}\psi_{{\bf k}, \sigma}
\ee
$F(|{\bf k}|)$ is a dimensionless form factor, which for all practical purposes, may be in ignored. In (\ref{coupling2}), an isotropic approximation to the lattice has been adopted. For the square symmetry of the lattice, $({\bf k \times q})$ in (\ref{coupling2}) is changed \cite{ASV2010} to 
\be
\label{lattcoup}
\big(\sin (k_xa) \sin (k_y'a) - \sin (k_ya) \sin (k_x'a)\big),
\ee
with $({\bf k-k}') = {\bf q}$. (\ref{lattcoup}) is essential in obtaining the variation in magnitude of the fermion scattering with angle on the Fermi-surface.

{\it Dissipation}:\\
In order to generate a contribution to the action due to dissipation, we can integrate over the fermions,  as in Fig. (\ref{Fig-damp-self-energy}-top) using the coupling vertex (\ref{coupling1}).
The intermediate state carries current due to the fermions  which dissipate in the limit $T \to 0$ due to impurity scattering. The dissipative term in the action in the long-wavelength limit is then
\be
\label{diss}
S_{diss} =  g_0^2 q^2 Im <JJ>(q=0, \omega) |\theta({\bf q}, \omega)|^2 \equiv i \frac{\alpha}{4\pi} \omega q^2  |\theta({\bf q}, \omega)|^2.
\ee
Here, the conductivity $\sigma = (1/\omega) Im <JJ>(q=0, \omega)$ for $T \to 0$ is used to define the parameter $\alpha$. This form of dissipation has the same physics and the same form as the derived by Caldeira and Leggett  \cite{CaldeiraLeggett} for a Josephson junction in contact with an ohmic bath. The parameter $\alpha$ introduced by them is equal to  $\frac{1}{4\pi^2}\sigma R_q$ , where $R_q = h/4e^2$ is the quantum of resistance. One should also include dissipation with the coupling (\ref{coupling2}) to the local angular momentum of the fermions. One again gets a similar form for the result.

\section{The Solution of the Dissipative Quantum XY Model}
The action of the (2+1)D quantum dissipative XY model for the angle $\theta({\bf r}, \tau)$ of fixed-length quantum rotors at space-imaginary time point $({\bf r}, \tau)$ is 
\begin{eqnarray}
\label{S1}
S &=&-K \sum_{\langle {\bf r, r}' \rangle} \int_0^{\beta} d \tau \cos(\theta_{{\bf r}, \tau} - \theta_{{\bf r}', \tau}) 
 + \frac {C}{2} \sum_{{\bf r}} \int_0^\beta d \tau \left( \frac{d \theta_{{\bf r}}}{d\tau}\right)^2  \nonumber \\
&+&  \frac{\alpha}{2} \sum_{\langle{\bf x, x}'\rangle} \int d \tau  d\tau' \frac {\pi^2}{\beta^2} \frac {\left[(\theta_{{\bf r}, \tau} - \theta_{{\bf r}', \tau})  -(\theta_{{\bf r}, \tau'} - \theta_{{\bf r}', \tau'}) \right]^2}{
\sin^2\left(\frac {\pi |\tau-\tau'|}{\beta}\right)}).
\label{eq:model}
\end{eqnarray}
 $\tau/2\pi$ is periodic in $\beta$, the inverse of temperature $1/(k_B T)$.  $\langle {\bf r, r}'\rangle$ denotes nearest neighbors. The first term is the spatial coupling term as in classical XY model. The second term is the kinetic energy where $C$ serves as the moment of inertia. The third term is the transformation of the dissipation of Eq. (\ref{diss}) to imaginary time and real space. 
 
\begin{figure}[tbh]
\centering
\includegraphics[width=0.6\columnwidth]{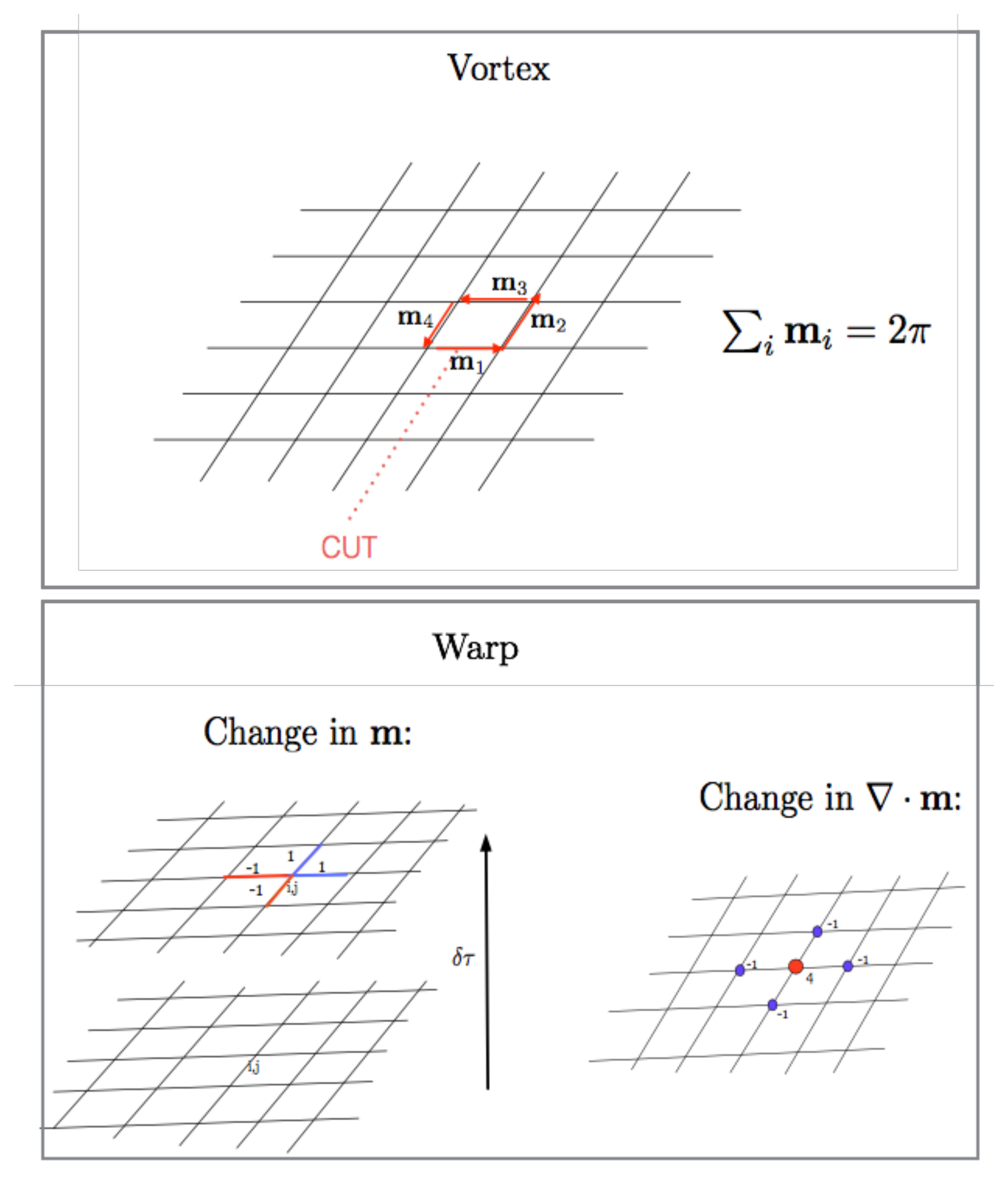}
\caption{The sketch at the top shows the configuration at a fixed time of the ${\bf m}$ field defining a vortex. At the bottom, the definition of a warp is sketched; it is a change of phase $\theta$ by $\pm 2\pi$ in a time step at a given point in space. This leads to a change in the field ${\bf m}$ in a time step, which is equivalent to the generation of a monopole with charge 4 surrounded by 4 anti-monopoles with charge -1 at the neighbors. In an anti-warp, the sign of the charges are reversed.}
\label{Fig_vorwarp}
\end{figure}

 In Ref. ~[\onlinecite{Aji-V-qcf1, Aji-V-qcf2}], it is shown that after making a Villain transformation  and integrating over the small oscillations or spin-waves, the action is expressed in terms of link variables which are differences of $\theta$'s at nearest neighbor sites, as shown in Fig.~(\ref{Fig_vorwarp}).
\be
\label{eq:m}
m_{{\bf r,r}'}(\tau, \tau') \equiv \theta({\bf r},\tau) - \theta({\bf r}',\tau').
\ee
Further
\be
{\bf m} = {\bf m}_{\ell} +{\bf m}_t
\ee
where ${\bf m}_{\ell}$, is the longitudinal (or curl-free) part  and ${\bf m}_t$ is the transverse (or divergence-free) part . The appearance of ${\bf m}_{\ell}$ is a novel feature of the quantum dissipative XY-model.
Now define
\be
\nabla \times {\bf m}_t ({\bf r},\tau) = \rho_v({\bf r}, \tau) \hat{\bf z},
\ee
so that $\rho_v({\bf r}, \tau)$ is the charge of the vortex at $({\bf r},\tau)$, and
\be
\frac{\partial {\hat{\nabla}}\cdot {\bf m}_{\ell}({\bf r}, \tau)}{\partial \tau} = \rho_w({\bf r}, \tau).
\ee
$\rho_w({\bf r}, \tau)$ is called the ``warp'' at $({\bf r},\tau)$.

Although a continuum description is being used for simplicity of writing, it is important to do the calculation so that the discrete nature of the $\rho_v, \rho_w$  fields is always obeyed. In the numerical implementation of (2+1)D discrete lattice,  given the two bonds per site $({\bf r})$, one may construct a vector field ${\bf m}_{{\bf r},\tau}$, whose components are the two directed link variables in the Cartesian directions:  
\be
m^x_{i,j,\tau} &=& \theta_{i+1,j,\tau}-\theta_{i,j,\tau},  \nonumber \\
m^y_{i,j,\tau} &=& \theta_{i,j+1,\tau}-\theta_{i,j,\tau},  
\label{eq:vecfieldnum}
\ee

A figure of the familiar vortex configuration for currents, and of the change in configuration of phases in successive time steps, actually seen in Monte-Carlo calculations is shown in 
Fig. (\ref{Fig_vorwarp}).

   In terms of the vortex and warp densities, the action of the model is (transformed from that shown in frequency-momentum space \cite{Aji-V-qcf1} to (imaginary) time and space, and dropping terms quadratic in $\rho_w$'s which are much less singular than the two leading terms kept, the action is 
\be
\label{topomodel}
S &=&\int d{\bf r}d{\bf r'}d\tau d\tau'  \Big( \frac{J}{2\pi}~ log{({\bf r}-{\bf r}')} \delta(\tau-\tau') \rho_v({\bf r},\tau)\rho_v({\bf r}',\tau') \\ \nonumber 
&+& \frac{\alpha}{4\pi}log{(\tau-\tau')}\delta({\bf r-r}') \rho_w({\bf r},\tau)\rho_w({\bf r}',\tau') + \frac{g}{\sqrt{|{\bf r}-{\bf r}'|^2 +v^2 |\tau-\tau'|^2}}\rho_w({\bf r},\tau)\rho_w({\bf r}',\tau')\Big).
\ee
Here the dimensionless parameters are $J = K_0\tau_c$  and $g = \sqrt{J/E_c}/4\pi,~ v^2/c^2 = KE_c, c =a/\tau_c$, and $\tau_c$ is the ultra-violet cut-off in the problems. 

The first term in (\ref{topomodel}) is the action of the {\it classical} vortices interacting with each other through logarithmic interactions in space but the interactions are local in time. The second term describes the warps interacting logarithmically in time but locally in space. The third term is just the action for a Coulomb field, which if present alone is known \cite{polyakov} not to cause a transition; it is marginally irrelevant in the present problem. The warp and the vortex variables in the first two terms are orthogonal. With just these two terms alone, the problem is exactly soluble. If the first term dominates, one expects a transition of the class of the classical Kosterlitz-Thouless transition through binding of vortex-anti-vortex pairs in space but there is nothing to order the vortices with respect to each other in time. If the second term dominates, there is a quantum transition to a phase with binding of warp-antiwarp pairs in time but nothing to order them with respect to each other in space.  Given the growth of correlations driven by either the density of isolated vortices or of isolated warps $\to 0$, the flow from one to the other and the ordered state is determined by the third term. This  leads to ordering at $T=0$ both in time and space to a state with symmetry of the 3D XY model. 
The transformation to the topological model above relies on a finite dissipation coefficient $\alpha$. With $\alpha=0$, the velocity field is divergence free and warps cannot be defined. In that case, the model is the same at $T=0$ as the 3D classical XY model. One of the results of the Monte-Carlo calculations is that at $\alpha \approx 0.1$, the transitions of the model (\ref{eq:model}) change from such a class to those being discussed here.

\begin{figure}[tbh]
\centering
\includegraphics[width=1.0\columnwidth]{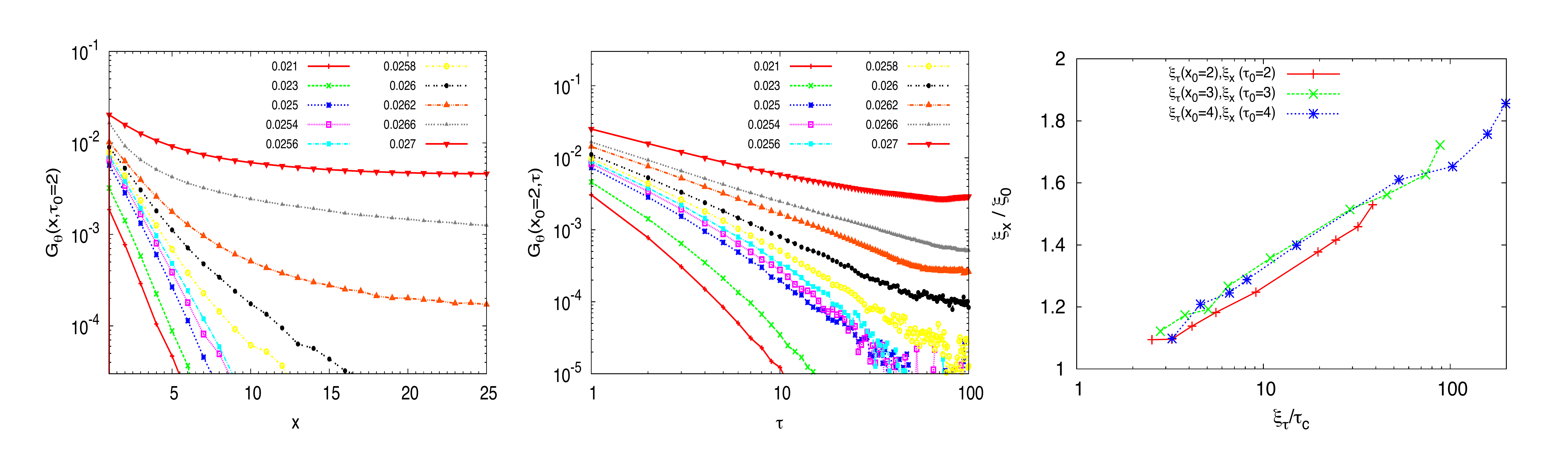}
\caption{The left panel shows the results of quantum-Monte-Carlo calculations of the spatial dependence of the correlation function of the quantum dissipative XY model for a fixed time, the middle the (imaginary) time-dependence of the correlation function for a fixed time, for various values of the parameter $\alpha$, through which the critical value $\alpha_c$ and the spatial and temporal dependence, described in the text are discovered. The right panel shows that the spatial correlation length diverges (within numerical uncertainty) as the logarithm of the temporal correlation length $\xi_{\tau}$. For details and many other calculations, see Ref. \cite{ZhuChenCMV2015}.}
\label{Fig-QCFXY}
\end{figure}

\subsection{Monte-Carlo Calculations} 
In Monte-Carlo calculations on the starting model (\ref{S1}), the  phase diagram of the model has been evaluated \cite{Stiansen-PRB2012, ZhuChenCMV2015}. (In these references, $\alpha$ is $1/4\pi^2$ times the $\alpha$ defined above.) Direct evidence
of vortices and warps though identifying configurations in space and time sketched in Fig. (\ref{Fig_vorwarp}) is obtained. One can conclude from the calculations \cite{ZhuChenCMV2015} of  their density and their correlations in time and space across the phase transitions that the representation of the model through the action for warps and vortices (\ref{topomodel} is faithful. There are three distinct phases found. The correlation functions for the order parameter 
\be
{\cal C}(r,\tau) \equiv <e^{i\theta(r, \tau)}e^{-i\theta(0,0)}>,
\ee
 are calculated at the transitions between them. 
In Fig. (\ref{Fig-QCFXY}), some results are shown for ${\cal C}(r,\tau)$ near the transition from a disordered to the fully ordered phase, which is driven by increasing the parameter $\alpha$, for $K/K_{\tau} \lesssim 4$. This is the relevant transition for the observed quantum-critical fluctuations in AFM's as well as the cuprates.  The results for ${\cal C}(r,\tau)$ for a constant $\tilde{K} \equiv {K_0/E_0}$ are expressible on the disordered side as,
\be
\label{corrfn}
{\cal C}({\bf r}, \tau) \approx  \chi_0 \frac{1}{\tau} e^{-\sqrt{\tau/\xi_{\tau}}} e^{-|r|/\xi_r}; ~~~
\xi_\tau = \tau_0 e^{-(\alpha_c/(|\alpha_c - \alpha)|)^{1/2}};~~~\xi_r/a \propto \log (\xi_\tau/\tau_0).
\ee
$a$ is the lattice constant in space and $\tau_0$ is the short-time cut-off, which is also calculated in terms of the parameters of the original model. These results are shown in Fig. (\ref{Fig-QCFXY} - right). 
If $\alpha$ is kept constant and the transition studied as a function of $\tilde{K}$, the correlation function retains its separable form but
\be
\label{corrfun2}
\xi_\tau = \tau_0 \big(\frac{\tilde{K}_c}{\tilde{K}-\tilde{K}_c}\big)^{\nu_{\tau}};~~~\xi_r/a \propto \log (\xi_\tau/\tau_0).
\ee
with $\nu_{\tau} \approx 1/2$. Note that the logarithmic relation between $\xi_r$ and $\xi_{\tau}$ is preserved. This form may be more relevant to the case of the antiferromagnets as well as the cuprates, where the transtion is most likely driven by the coupling constants in the potential and kinetic energies, rather than in the dissipation parameter.-

There are three extra-ordinary features in (\ref{corrfn},\ref{corrfun2}): (1) The correlation function is separable in time and space, unlike in the LGW class of quantum theories. (2) The spatial correlation length $\xi_r$ is proportional to the logarithm temporal correlation time $\xi_{\tau}$. One might say that the dynamical critical exponent $z =\infty$, but this can be misleading. Besides, as explained below $z$ itself is a flowing scale-dependent variable.  (3) At criticality, the correlation function is $\propto \tau^{-1}$;  this Fourier transforms to give the imaginary part $\propto \tanh (\omega/2k_BT)$. The last is precisely the ansatz \cite{CMV-MFL} for critical phenomena on which the {\it marginal fermi-liquid} is based. But unlike the assumption made in that ansatz, there is a diverging spatial correlation length, though its divergence is extremely slow compared to the divergence of the temporal correlation length. 

The $\tau$-dependence in (\ref{corrfn}) can only be Fourier transformed numerically \cite{ZhuChenCMV2015}, because of the square-root in the exponent. If it is changed to linear in $(-\tau/\xi_{\tau})$, the imaginary part of the correlation function for AFM quantum-criticality is
 \be
\label{chi-tr}
{\cal C}''({\bf q}, E, T) &=& - \chi_0  \tanh\left(\frac{E}{\sqrt{(2T)^2 + \xi_{\tau}^{-2}}}\right) \frac{1}{|{\bf q}|^2 + \xi_r^{-2}}
\ee 
It should be remembered that (\ref{chi-tr}) is only valid in the quantum-critical regime. For example, one can use this form for the correlation function with a temperature independent $\xi_r$ only for T much less than the upper energy cut-off $\tau_c^{-1}$. Also, one must be in the regime of 2 D spatial fluctuations.

These results are quite different from those based on Landau-Ginzburg-Wilson type of theories or the extensions of classical dynamical critical phenomena to the quantum regime, pioneered by Moriya \cite{Moriya-book}, Hertz \cite{Hertz}, Beal-Monod and Maki \cite{Beal-Monod} and by others \cite{Millis-qcf, Belitz-RMP}. In such theories, critical modes are soft with a diverging amplitude at low energies. In contrast, the distribution in frequency of the spectral weight in the correlation function (\ref{corrfn}) remains unchanged as the critical point is approached. Since $lim(T \to 0) \tanh(\omega/2T) \to Sign(\omega)$, only the part for $\omega << 2T$ increases from linear in $\omega/T$ to a constant as $T \to 0$, with a jump discontinuity in going across 0 in the real axis. This, as well as the logarithmically slow 
increase of the spatial correlation length, are essential in deriving the observed scattering rates, temperature and frequency dependence in transport and the weak divergences in thermodynamic properties. This will be further elaborated below.

 According to the discussion above on mapping to the XY model, the anti-ferromagnetic correlations for 2D quantum-critical fluctuations for incommensurate uni-axial correlations and commensurate or incommensurate planar correlations is given by Eq. (\ref{chi-tr}), with ${\bf q}$ replaced by $({\bf q-Q})$, where the correlations as a function of ${\bf q}$ peak at ${\bf Q}$. 

\subsection{Renormalization Group Calculations}
Confidence is gained on the principal results from the Monte-Carlo calculations on the model (\ref{S1}) by reproducing (most of ) them \cite{Hou-CMV-RG}  in leading order renormalization group calculations, on the equivalent model (\ref{topomodel}). The most interesting result is that the flow towards criticality of $\alpha$ drives through the flow of the warp fugacity to a critical flow of $v$ or equivalently of the dynamical critical exponent $z$. This in turn drives the flow of the fugacity of the vortices such that the results for the correlation function (\ref{corrfn}) are obtained.

\section{Applications and Tests of Theory for Response Functions, Thermodynamics and Transport}
\begin{figure}[b]
\centering
\includegraphics[width=1.0\columnwidth]{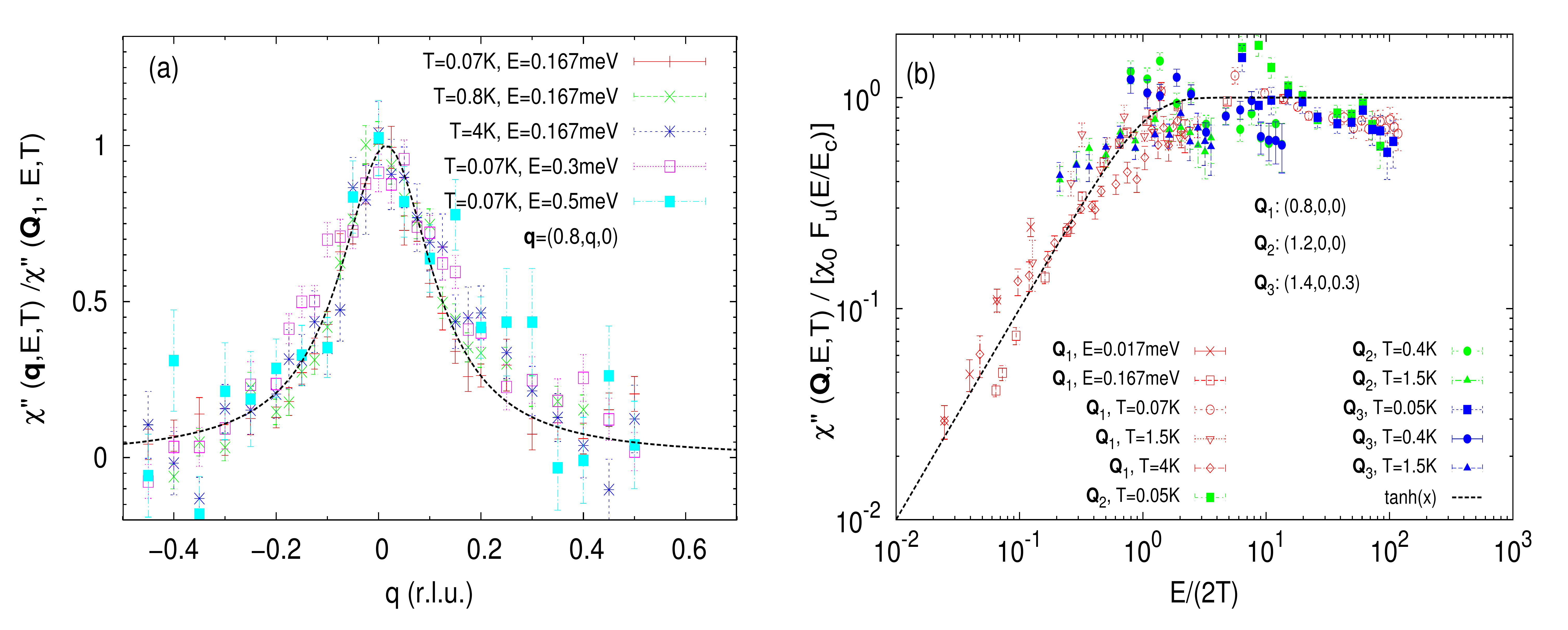}
\caption{Left: $\chi '' ({\bf q}, E, T)$ as functions of ${\bf q}$ for a $q$-scan across ${\bf Q}$=(0.8,0,0), at various fixed $E$ and $T$ for CeCu$_{5.9}$Au$_{0.1}$. The fitting curve is Lorentzian $1/[(q-q_c)^2/\kappa_q^2+1]$ with $\kappa_q$=0.11 r.l.u. $\approx$ 0.13 \AA$^{-1}$ (considering $b=5.1$\AA). Right: $\chi ''({\bf Q}, E, T)$ as functions of $E/(2T)$ for various constant-$E$ or $T$ scans. The solid line is 
$\tanh(E/2T)$. Original data is taken from \cite{Schroder1, Schroder2}. For details of the fits and the re-plotting of the data, see \cite{CMVZhuSchroeder2015}.}
\label{fig:chiq}
\end{figure}

The most stringent test of the theory is through the measurement of $\chi( {\bf q}, E,T)$, from which most other properties can be derived. The results with the neutron scattering measurements in the heavy fermion compound CeCu$_{5.9}$Au$_{0.1}$ at the various indicated frequencies and temperatures are presented for a momentum scan across the AFM vector are shown in Fig. (\ref{fig:chiq}-left). The scaling of the frequency and temperature are are presented in Fig. (\ref{fig:chiq}-right). As shown, the distribution in $q$ about the maximum fits a Lorentzian, with a width, which is consistent with being frequency and temperature independent to within the error bars, in the range of a factor of 3 in frequency and 50 in temperature, over which it has been measured. This is consistent with the theoretical result that the q and the $E, T$-dependence are separable.  The Moriya-Hertz type theory would have the inverse correlation length (the half-width in Fig. (\ref{fig:chiq}) proportional to $(T, \omega)^{-1/2}$.
The frequency and temperature dependence are also consistent with the form expected at criticality, i.e. $\propto \tanh(E/2T)$, when account is taken for the fact that the measurements extend across the fermi-energy of the compound, as explained in \cite{CMVZhuSchroeder2015}. In Ref. \cite{CMVZhuSchroeder2015}, comparison with the theoretical expectations is also presented for data in BaFe$_{1.85}$Co$_{0.15}$As$_2$ measured in \cite{Inosov2010}. Some recent systematic measurements on Ba(Fe$_{0.957}$Cu$_{0.043})_2$As$_2$, \cite{MGKim2015} are consistent with a $\tanh(\omega/2T)$ dependece of the peak of the q-dependent correlation function over a wide range of $\omega$ and $T$. But the width of the q-dependence shows a complicated dependence which may be fitted to a constant at T $\gtrsim 150 K$ crossing over to a divergent behavior at temperatures below about 20 K. It is known that large 3D correlations develop in this compound at low temperatures.

It is amusing to note that early measurements \cite{HaydenPRL1991, KeimerPRL1991, BourgesPR1991} of cuprates near the AFM quantum-critical region at very low doping, in Fig. (\ref{Fig:CommonPhDia}) found AFM correlation lengths, which were temperature independent and with a frequency dependence consistent with $\propto tanh(\omega/2T)$, above a temperature below which spin-glass type order sets in. For larger dopings, the AFM correlation length rapidly becomes of the order of a lattice constant \cite{Bourges-Balatsky-AFM}, showing that AFM correlations can be disregarded for the region of quantum-critical phenomena near optimal doping.

\subsection{Self-energy, Normal State Scattering Rate, Resistivity and Other Anomalies}

It is useful to start with the exact relation \cite{Nozieres-book} of the one-particle self-energy $\Sigma({\bf p}, \epsilon)$ to the irreducible vertex in the particle-hole channel, $I_{irr}^{ph}({\bf p,p',q}; \epsilon, \epsilon', \nu)$ and the exact single-particle Greens' function $G({\bf p'}, \epsilon')$ in the normal state. This is shown, for the normal state in the left of Fig. (\ref{SE}). The vertex is irreducible in the particle-hole channel with total momentum-energy (${\bf q}, \nu)$ and it is assumed, as usual, that it is regular in the limit $({\bf q}, \nu) \to 0$ in this channel, which alone is needed in the self-energy calculations. When the vertex depends on $(\omega, \omega')$ through only $(\omega - \omega')$, Fig. (\ref{SE}) is exactly equivalent to the skeleton diagram (\ref{Fig-damp-self-energy}-a). The associated integral equation for the self-energy given below includes all  "vertex corrections"  and self-energy insertions of the perturbative calculations. 

We are interested only in the singular contributions to the self-energy due to exchange of the collective fluctuations, specified  by Eq. (\ref{chi-tr}) of the paper.  In this case, the irreducible vertex in Fig. (\ref{Fig-damp-self-energy}-a) 
\be
\label{vertex}
I_{irr}^{ph}({\bf p,p',0}, \epsilon, \epsilon', 0) = |g({\bf p},{\bf p'})|^2 C({\bf p,p'}, \epsilon-\epsilon').
\ee

Given the momentum dependence of $C({\bf p-p}', \epsilon-\epsilon')$ of the form (\ref{chi-tr}) and the dependence of  $g({\bf p,p'})$ of either the form (\ref{coupling1}) or (\ref{coupling2}), it is safe to begin by taking $g({\bf p,p'})$ to be a constant $g_0$ for calculating self-energy in the normal state which is required to have the full symmetry of the lattice. (This point is discussed further in Sec. VI below.) 
Following the procedure described in Ref.\cite{AGD}-sec-23.1, the self-energy is given by
\be
\label{s-1}
\Sigma({\bf p}, \epsilon) &=& \frac{g_0^2}{\pi (2\pi)^d} \int d{\bf p'} \int_{-\infty}^{\infty} d\omega' \int_{-\infty}^{\infty} d\epsilon_1 \\ \nonumber
& \times & \frac{Im G_R({\bf p'}, \epsilon_1) Im C_R({\bf p-p'}, \omega')}{\omega' + \epsilon_1 -\omega - i\delta} \big(\tanh \frac{\epsilon_1}{2T}+ \coth \frac{\omega'}{2T}\big)
\ee
$C_R$ is the retarded fluctuation propagator and $G_R$ is the retarded one-particle propagator. We can follow the steps given in Ref. (\onlinecite{AGD})-sec-23.1 for evaluating the integrals in (\ref{s-1}), except that we do not assume that the imaginary part of the self-energy is insignificant as for phonons, or assume the Migdal approximation. But as in Ref. (\onlinecite{AGD}), we  assume that given the form of $C$, we expect the self-energy to be momentum independent. This is expected, of-course if $C$ were to be momentum independent, but as we will see, it is true also if $C$ is separable in momentum and frequency, as in Eq. (17) in the paper. Then $G({\bf p}, \epsilon)$ is given in terms of the non-interacting band-energy $\xi_{\bf p}$ and the self-energy which is to be solved for by
\be
G({\bf p}, \epsilon) = \frac{1}{\epsilon - \xi_{\bf p} - \Sigma(\epsilon)}.
\ee
Using this, we get from Eq. (\ref{s-1}) that the imaginary part of the self-energy is
\be
\label{s-2}
Im \Sigma_R({\bf p}, \epsilon) &= &\frac{\pi g_0^2}{(2\pi)^2}\frac{m}{p_F}\int_0^{k_c} dk  G(k) \int_{-\infty}^{\infty} d\omega C_R(\omega) \big(\tanh \frac{\epsilon+\omega}{2T}+ \coth \frac{\omega}{2T}\big) \\ \nonumber
& \times &\big({\cal{T}}^{-1}(\epsilon + \omega, \xi_{|{\bf p}|+ k}) - {\cal{T}}^{-1} (\epsilon + \omega, \xi_{|{\bf p}| - k})\big).
\ee
The integrations have been performed using  the separable form of the fluctuation propagator given by Eq. (\ref{chi-tr}).
$k_c$ is an upper-cutoff for the magnitude of momentum transfer, which is the zone-boundary, and  
\be
\label{T}
{\cal{T}}^{-1}(x,y) = \arctan\big( \frac{x- Re\Sigma(x) - y}{Im \Sigma(x)}\big); ~~ \xi_{|{\bf p}| \pm k} = \big((|{\bf p}| \pm k)^2 -p_F^2\big)/2m.
\ee
We have also specialized to 2d (although that is not necessary) and dropped a factor in the Jacobian for converting from momentum to energy integrals, which becomes important only in the region of forward scattering which is unimportant in the integral.
We expect the self-energies to be in the same scale as $\epsilon$ for $\epsilon \gtrsim T$ and on the scale of $T$ for $\epsilon \lesssim T$, i.e. smaller than the upper range $\xi(k_c)$ of the $\xi$'s. (The calculation below does not change if there are logarithmic correction to $Re \Sigma (\epsilon)$). Given the range of the $k$-integral, the restrictions on the 
$\omega$-integral from the ${\cal{T}}$ factors is over the band-width $\xi(k_c)\pm \Sigma(\epsilon)$ corrections. The corrections due to 
$\Sigma(\epsilon)$ are un-important for $\epsilon$ of interest because the range of $\omega$ integration is actually
 limited by the thermal factors in (\ref{s-2}) to the much smaller energies of $O\big(max(\epsilon, T)\big)$. 
The upper limit on the integral over $k$ can therefore be done easily over its entire range. We are left only with the $\omega$ integral. In the quantum-critical regime, the temporal corelation length in Eq. (17) of the paper $\xi_{\tau} << T$, so that $Im C_R(\omega) = -\chi_0 \tanh{(\omega/2T)}$. In this regime the self-energy is then given by
\be
\label{self-e}
Im \Sigma_R({\bf p}, \epsilon) &=& \overline{g}_0^2 N(0) \chi_0 max(|\epsilon|, T), ~\text{for}~ max(|\epsilon|, T) \lesssim \omega_c, \\ \nonumber
&= & \overline{g_0}^2 N(0) \chi_0 \omega_c, ~\text{for} ~max(|\epsilon|, T) \gtrsim \omega_c
\ee
$\overline{g_0}$ includes numerical corrections of O(1) to $g_0$, which depend on details of the band-structure. 

For the regime, $\xi_{\tau}^{-1} >> T$, the integral over $\omega$ is cut-off by $\xi_{\tau}^{-1}$ and the contribution to self-energy becomes $\omega^2 \xi_{\tau}$  which  vanishes as one deviates far from the critical point. The normal non-singular Fermi-liquid scattering which is always present takes over.

These results are similar in functional form to the perturbative results. That they are true more generally was stated without proof in Ref. (\cite{Kotliar-epl1991}) and the relations of the irreducible vertex to the complete vertex and to density-density correlations in the hydrodynamic regime were derived in Ref. (\cite{Shekhter-V-Hydro}). Following the microscopic theory of the fluctuations and the derivation of the coupling of the fermions to the fluctuations,  the same form of the results are shown to be observed for collective fluctuations which are separable in their momentum and frequency dependence, as for  as local ($q$-independent) fluctuations which were assumed in the phenomenology \cite{CMV-MFL}.  

In the above, we have used the coupling (\ref{coupling2})
for the isotropic approximation to the lattice. If the more appropriate coupling (\ref{lattcoup}) is used, an anisotropy of a factor of O(1) in the linear in $(\omega,T)$ self-energy is found \cite{ZhuVPRL2008} with a maximum in the $(\pi,0)$ and a minimum in the $(\pi,\pi)$ directions, so that the single-particle relaxation rate is of the form $\propto (1+ \alpha \cos(4\theta)$, where $\theta$ is the angle in the plane measured with respect to the crystalline axes, and $\alpha < 1$. This is also what is found in the analysis of anisotropy in the in-plane transport scattering rate found in detailed measurements using variations in resistivity with direction of magnetic fields \cite{Hussey2006, Hussey2012}.

The important predictions  from Eq. (\ref{self-e}), for cuprates, where the fluctuations are peaked near $Q=0$ is that the single-particle scattering rate is linear in $\omega$ and nearly independent of momenta ${\bf k}$ perpendicular to the Fermi-surface and varying only by factors of about 2 along the Fermi-surface.. This was verified for cuprates \cite{VallaPRL2000, KaminskiPR2005, Bok_ScienceADV, AbrahamsV-PNAS, ZhuVPRL2008} through angle resolved photoemission spectrscopy (ARPES) \cite{ZX-Rev}. 

For a momentum-independent self-energy, there is no backward scattering vertex correction for current transport.  (For angular dependent self-energy of the form mentioned above, the resistivity has the same angular dependence as the self-energy given a corresponding velocity asymmetry). This was used in \cite{Abrahams-V-Hall} to derive the resistivity proportional to $T$ in a solution of the Boltzmann equation including the full collision operator. The same result is obtained \cite{Shekhter-V-Hydro} more formally by deriving the density-density correlation for a marginal Fermi-liquid of the conserving form with a diffusion constant proportional to $Im \Sigma$. Using the relation between the density-density and the current-current correlations, the result for the resistivity $\propto T$ is again obtained. A small correction between the anisotropy of the single-particle scattering rate and the transport scattering rate should however occur. 

The detailed measurements of the scattering rate \cite{Hussey2006, Hussey2012} have revealed in addition to the above a Fermi-liquid contribution proportional to $T^2$. This is not surprising. The singularities leading to a marginal fermi-liquid are the leading contributions to the scattering rate but they do not eliminate the normal Fermi-liquid processes. In fact in the derivation \cite{Shekhter-V-Hydro} of the long wave-length structure factor for the marginal Fermi-liquid, Fermi-liquid renormalizations modify the results quantitatively.

One can turn to the exact expression for the entropy in terms of the single-particle Green's function given as Eq. (19.27) in Ref.\cite{AGD} to find that with (\ref{self-e}), the specific heat has a singular contribution $\propto T \ln T$. In cuprates, it is hard to deduce the electronic specific heat at temperatures above $T_c$ accurately, because of the much larger lattice specific heat. Thermopower, which is the entropy per carrier, has however been measured and is indeed $\propto T \ln T$ \cite{Taillefer-Thermopower2009}. The resistivity and the entropy/thermopower in the region of AFM quantum-criticality of the Fe-compounds and of the heavy fermions has already been mentioned.  Forward scattering due to impurities with elastic scattering rate varying on the fermi-surface due to variations in the local fermi-velocity \cite{Abrahams-V-Hall} contributes importantly to the measured Hall angle \cite{Ong-HallAngle}. However, the
contribution to the scattering rate varying as $T^2$ has been calculated \cite{Hussey2012} to give a larger contribution to the Hall angle than to the resistivity, leading also to a contribution to the anomaly in the Hall angle.

The optical conductivity at frequencies below about 1500 cm$^{-1}$ is calculated \cite{Abrahams1996} to scale as $\omega^{-1}$, with logarithmic corrections due to the logarithmically diverging effective mass, which vanish for $\omega \to 0$, as required by a Ward identity due to charge conservation. There has been some discussion of the apparent $\omega^{-2/3}$ form for the frequency dependent conductivity \cite{VanDerMarel2009} in an intermediate range of frequencies, between about 2000-4000 cm$^{-1}$. Such crossovers are required due to the cut-off $\omega_c \approx 4000 cm^{-1}$ in the fluctuation spectra. This leads to a saturation in the imaginary part of the self-energy above $\omega_c$.  This saturation must be accompanied by a corresponding change in the real part of the self-energy. Direct measurements of the real and imaginary part of the self-energy by ARPES (see Fig.(\ref{fig:Arpes}) below) spread over from about half  to about twice the cut-off show this behavior. Calculation of the optical conductivity using self-energies of similar form \cite{Abrahams1996} do show crossovers consistent with the observations. 

The anomalous thermodynamic and transport results have not been obtained from AFM quantum criticality, or indeed quantum-criticality of any other order parameter, either at $Q=0$ or finite ${\bf Q}$, with correlations of the Moriya-Hertz form. The frequency (temperature) dependence of the normal self-energy (for AFM or Charge density wave criticality) given by such correlations is angle-dependent, being non-Fermi-liquid like only in region near points on the Fermi-surface which are connected by the AFM wave-vector. The width of such regions decreases for increasing AFM correlation length. The correlation length measured for various $\delta$ in YBa$_2$Cu$_3$O$_{6+\delta}$ decreases to about a lattice constant near optimal doping \cite{Bourges-Balatsky-AFM}. 
 Elaborate dynamical mean-field calculation on the Hubbard model for various doping \cite{Gull-Millis2015, Gull-Millis2014} bear no relation to the measured frequency and temperature dependence of scattering rates by ARPES \cite{Johnson2000, KaminskiPR2005, Bok_ScienceADV}. Naturally, no calculations, with such ideas has yielded the linear in T resistivity, or the observed frequency dependence of the conductivity. 
 Nor have such results been obtained in any systematic calculation using the ideas of resonating valence bonds \cite{Anderson-book}.

\section{Applications to Superconductivity}

In this section, the unique features in superconductivity induced by exchange of fluctuations of the XY model are highlighted. Highly accurate angle-resolved single-particle spectroscopy has been used to test the theory. As many calculations attest \cite{ScalapinoRMP}, AFM fluctuations of the Moriya-Hertz form and on Hubbard model \cite{Tremblay2013, Gull-Millis2015} do give d-wave superconductivity with the right scale of $T_c$ if the antiferromagnetic correlation lengths are long enough \cite{NishiymaMV2013}. Such ideas work perfectly well in 3D- heavy fermion supercondutors
near their AFM quantum-criticality \cite{Stockert2011} for which they were originally proposed \cite{MiyakeSV1986, Scalapino1986}.

It appears inescapable, on looking at the phase diagram of the cuprates, the Fe-based compounds, and the heavy fermions, that in each case, superconductivity is promoted by the same fluctuations which lead to the anomalous properties in their quantum critical region. On one side of this region, the occurrence of the ordered phase due to the condensation of such fluctuations at finite temperature produces a low energy depletion of such fluctuations. On the other side of this region, the cross-over to a Fermi-liquid region again cuts off the low energy singularities of the fluctuations. This naturally leads to a decrease of the superconducting transition as one moves away from the critical region. The connection of superconductivity to normal state properties can be quantitatively established by analysis of angle-resolved photoemission in the normal and the superconducting state. 


The three most important properties in relation to superconductivity are, (A) the symmetry of Cooper pairs induced by the fluctuations and their coupling to Fermions, (B) the magnitude of the coupling constants obtained by appropriate averages of the coupling vertex of the fermions and the fluctuations and (C) the form of the energy dependence of the fluctuations and their upper cut-off $\omega_c$. I discuss (A) immediately below. (B) and (C) deduced from ARPES experiments, are discussed next. \\

\subsubsection{The symmetry of Cooper pairs induced by the fluctuations}

A basic result about the symmetry of superconductivity is that, s-wave pairing is induced when the scattering of fermions is nearly isotropic in the angle in momentum space through which they are scattered by the fluctuations,  p-wave pairing when the scattering is peaked at  $\pm \pi $, and d-wave pairing when it is peaked at $\pm \pi/2$, etc. \cite{MiyakeSV1986, Scalapino1986, VrevROP2012}. This result, for a nearly isotropic fermi-surface, has its obvious generalization to fermi-surfaces in actual lattices in terms of their irreducible representations.

\begin{figure}[t]
\centering
\includegraphics[width=0.8\columnwidth]{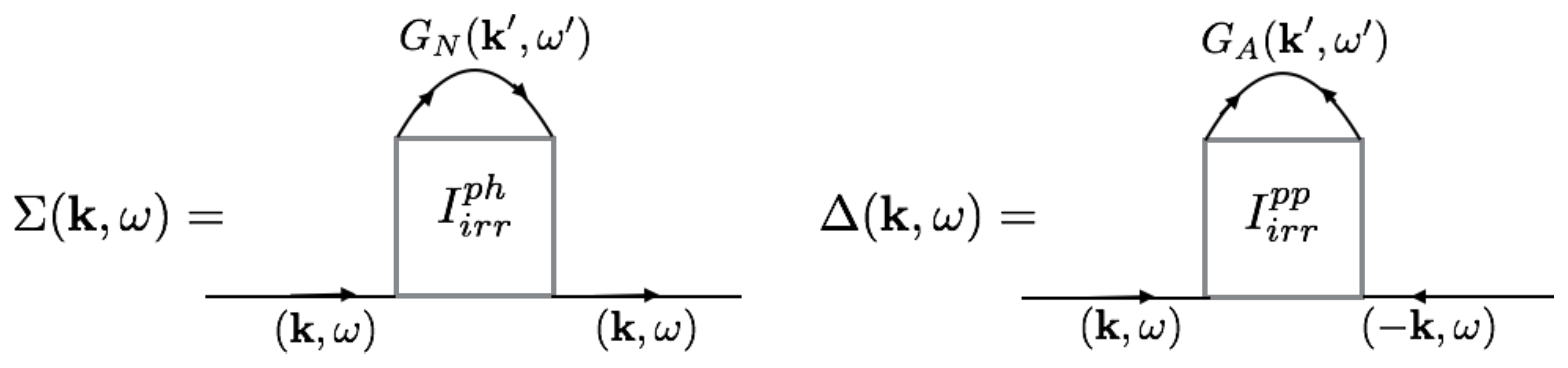}
\caption{Exact representation of the normal self-energy and the anomalous or pairing self-energy in the superconducting state in terms of the irreducible vertices in the particle-hole and particle-particle channels, respectively, and the corresponding parts of the exact single-particle Green's functions.}
\label{SE}
\end{figure}

We briefly discuss here the special features of d-wave superconductivity with fluctuations of the XY model and their coupling to fermions. For details, please see Refs.( \cite{ASV2010, Bok_ScienceADV}- (supplement)).  Consider the expression \cite{AGD, Nozieres-book} for the normal self-energy $\Sigma({\bf k}, \omega)$ and the pairing self-energy given in Fig. (\ref{SE}) in terms of, respectively, the irreducible particle-hole vertex $I_{irr}^{ph}\big({\bf k,k}',({\bf q}=\Omega=0), \omega,\omega')$ and the irreducible particle-particle vertex $I_{irr}^{pp}\big({\bf k,k}',({\bf q}=\Omega=0), \omega,\omega')$. These expressions are exact when the dependence on $\omega, \omega'$ is of the form $(\omega-\omega')$. In that case, the self-energy is equivalently given by the skeleton diagrams of Fig. (\ref{Fig-damp-self-energy}-a and b) with 
\be
\label{I-vertex}
\Big(|g({\bf k,k'})|^2 , g({\bf k}, {\bf k}')g(-{\bf k}, -{\bf k}')\Big)\cal{F}({\bf k, k'}, \omega, \omega')  \equiv  {\mathcal{I}}({\bf k, k', q=0}, (\omega-\omega'),\Omega = 0)),
\ee
 In (\ref{I-vertex}), $\mathcal{I} = I_{irr}^{ph} \tau_3 \tau_3 +I_{irr}^{pp} \tau_1\tau_1$.  The particle-hole irreducible vertex is in the $\tau_3 \tau_3$-channel and the particle-particle irreducible vertex is in the $\tau_1 \tau_1$ channel in the Gorkov-Nambu representation of the exact single-particle Green's functions in the superconducting state:
 \be
 \label{matrixGreen}
{\widehat G}({\bf k}, \omega) &=& \frac{W({\bf k}, \omega)\tau_0 +Y({\bf k}, \omega)\tau_3 +\phi({\bf k},\omega)\tau_1 }
 {W^2({\bf k}, \omega) - Y^2({\bf k}, \omega) -\phi^2({\bf k}, \omega)}
\ee
A further requirements is that $\mathcal{I}({\bf k, k'}, \omega, \omega'; {\bf q}, \Omega)$ should have a non-singular limit of the zero energy and momentum transfer in the irreducible channel, i.e for ${\bf q} \to 0, \Omega \to 0$. Eqs. in Fig. (\ref{SE}) are equivalent to

\be
 \label{self-energy-exact}
 {\widehat \Sigma}({\bf k}, \omega) = \int d\omega' ~Tr \sum_{{\bf k'}} \mathcal{I}({\bf k, k'}, \omega-\omega'; {\bf q} \to 0, \Omega \to 0) {\widehat G}({\bf k}', \omega').
 \ee
In the (skeleton) diagram, Fig. (\ref{Fig-damp-self-energy}-a), the intermediate propagator at $({\bf k}', \omega')$ is that of a single-particle state projected to the full symmetry of the lattice.  The summation over ${\bf k}'$ on evaluating the $\Sigma({\bf k}, \omega)$ 
 then gives the projection to identity of the product 
$|g({\bf k}, {\bf k}')|^2 Im {\cal F}({\bf k}, {\bf k}', \omega)$. Given the form of $\cal{F}$ and the cancellation of its singularity as a function of $({\bf k - k}')$ with the dependence on magnitude $|k-k'|^2$ in $|g({\bf k}, {\bf k}')|^2$, this projection is given only by the angular dependences in $|g({\bf k}, {\bf k}'|^2)$. Given Eq. (\ref{coupling2}),
 \be
 \label{coupling3}
|g(\hat{{\bf k}}, \hat{{\bf k}}')|^2 = -g(\hat{{\bf k}}, \hat{{\bf k}}')g(-\hat{{\bf k}}, -\hat{{\bf k}}') =  \left[1- (\cos 2\theta \cos 2\theta'+ \sin 2\theta \sin2\theta')\right],
 \ee
It then follows that only the first term in Eq.\ (\ref{coupling3}) then contributes on integration over $\theta'$. One therefore finds that $\Sigma({\bf k}, \omega)$ is isotropic. This result changes for a square lattice if the velocity $v({\bf k})$ is anisotropic and gives beside the isotropic contribution, a leading contribution $\propto \cos 4\theta({{\bf k}})$. 

Consider $\Delta(\theta, \omega)$ given by Fig. (\ref{Fig-damp-self-energy}-b). This is non-zero only in the superconducting state, because the intermediate state is itself proportional to $\phi(\theta', \omega')$. The intermediate state is the anomalous or $\tau_1$ part of $G({\bf k}', \omega')$, which has the symmetry of pairing, i.e. of $\phi({\bf k}', \omega) \propto \cos (2 \theta_{{\bf k}'})$. It is easy to see that couplings of the form (\ref{coupling1}) cannot contribute to such a pairing. Only the second term in Eq.\ (\ref{coupling3}) contributes on integration over $\theta'$, so that $\phi(\theta, \omega) \propto \cos(2\theta)$. From Eq. (\ref{coupling3}), it also follows that this part of the vertex is attractive while the s-wave part is repulsive in the pairing channel.


\subsubsection{Experimental Tests for Superconductivity in Cuprates}
\begin{figure}[tbh]
\centering
\includegraphics[width=1.0\columnwidth]{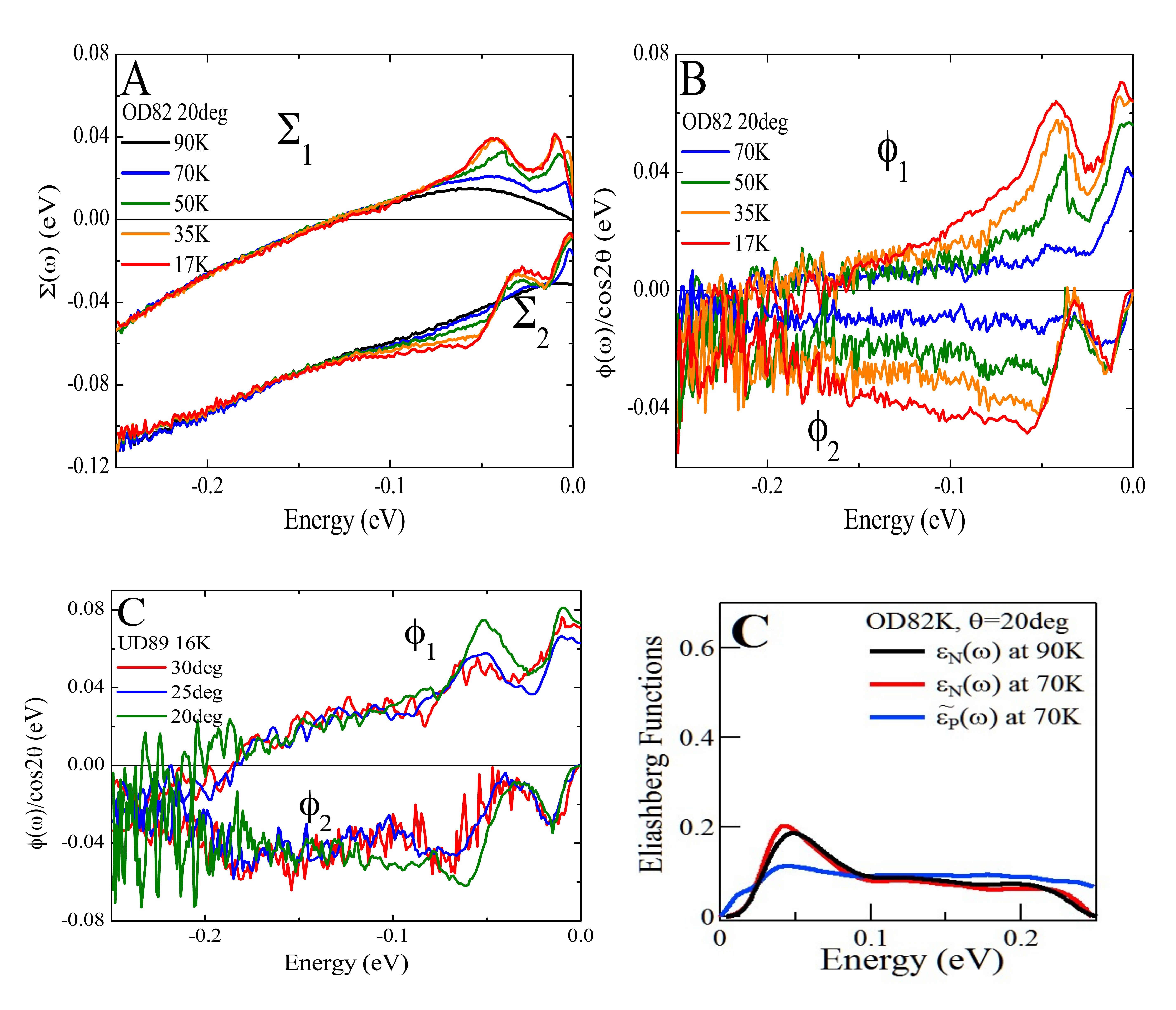}
\caption{The normal $\Sigma({\bf k}, \omega)$ and pairing self-energy $\phi({\bf k}, \omega)$ and the effective interaction vertices derived directly from the high resolution angle-resolved photo-emission data in samples of Bi2212. Top left shows $\Sigma({\bf k}, \omega)$ of a sample with $T_c =82 K$ at $25^\circ$ from the diagonal as a function of temperature. Besides the superconductivity induced features at energies below about 3 $\Delta$, $\Sigma({\bf k}, \omega)$ remains linear in $\omega$ and nearly independent of ${\bf k}$. $\phi({\bf k}, \omega)$  increases as $T$ decreases below $T_c$ and is $\propto cos(2\theta(\hat{{\bf k}}))$. Near $T_c$, the effective interactions in the d-wave channel ${\tilde{\cal E}_P}(\omega)$ has the $\omega$-dependence consistent with the quantum-critical fluctuations of the quantum XY model for loop-current fluctuations and within experimental uncertainty is the same as the (repulsive) interactions
in the full symmetry of the lattice ${\cal E}_N(\omega)$ , except for a weak feature at about 50 meV in the latter.}
\label{fig:Arpes}
\end{figure}

The quantitative analysis by McMillan and Rowell \cite{McMillan-Rowell} (MR) of very precise tunneling spectroscopy,  using Eliashberg generalization of BCS theory \cite{Eliashberg1960, SSW}, decisively confirmed that the exchange of phonons by the Fermions is responsible for the conventional s-wave superconductivity in metals such as Pb. Tunneling experiments integrate over the momentum dependence of the many body effects. This is sufficient for s-wave superconductors because the normal and the Cooper pairing interaction energies (self-energies), shown in Fig.(\ref{Fig-damp-self-energy}-(a,b)), have the full symmetry of the lattice. Since for cuprates the dependence on ${\bf k}$ of the pairing self-energy  $\phi({\bf k}, \omega)$ has $B_{1g}$ or $d_{x^2-y^2}$ symmetry, both the momentum and the frequency dependence of the interactions is necessary to decipher the fundamental physics.
The much more technical ARPES experiments and a much more detailed analysis are then necessary.  

Recently the single-particle self-energies in the pairing and the full lattice symmetry have been deduced directly from the high resolution laser based ARPES data \cite{Bok_ScienceADV} on two samples of Bi2212 in range of angles from the diagonal in the Brillouin zone to 25$^\circ$ from it.  Some results for the normal and pairing self-energy are shown in Fig (\ref{fig:Arpes}). They are used to deduce the magnitude and the frequency dependence of the effective interactions  both in the full symmetry of the lattice ${\cal E}_N(|{\bf k}|, \omega)$, and in the pairing symmetry ${{\cal E}}_P(|{\bf k}|, \omega)$. The latter are also shown in Fig (\ref{fig:Arpes}). These are the so-called Eliashberg functions, which are identical for s-wave superconductors, and often denoted by $\alpha^2F(\omega)$. The experimental results and the analysis, have been fully described elsewhere \cite{Bok_ScienceADV}. It is also shown there that the procedure for deducing these fluctuations is correct even when the high energy cut-off is similar to the electronic band-width. In other words, no Migdal approximation or neglect of vertex corrections is necessary.

The principal conclusions are that   near $T_c$, the attractive interactions ${\tilde{\cal E}}_P(|{\bf k}|, \omega) \equiv {\cal E}_P(|{\bf k}|, \omega)/\cos(2\theta_{\bf k})$ are, within the experimental uncertainty, identical to the repulsive interactions ${\cal E}_N(|{\bf k}|, \omega)$, except for a weak repulsive part near about 50 meV, present only in the latter. Both are independent of $|{\bf k}|$ and their major part is consistent with the quantum-critical fluctuations of the form given by Eq. (\ref{chi-tr}), and with the coupling functions with properties consistent with (\ref{coupling3}). The dimensionless coupling constant, which determines the normal scattering rate and resistivity is weak, $\approx$ 0.15, but the upper cut-off of scatterers $\omega_c$  is large, of O(0.4 eV). For this form of the quantum-critical fluctuations, the coupling constants for superconductivity are enhanced by $O(log(\omega_c/T_c))$  with respect to normal scattering coupling constant due to its frequency independence from the cut-off down to $T_c$. The coupling constants and the cut-off give a reasonable estimate of $T_c$. To within factors of O(2), such coupling constants and cut-offs were estimated from microscopic theory \cite{ASV2010}.

The measured self-energies have also been compared \cite{Bok_ScienceADV} to calculations based on the measured AFM fluctuations (in LSCO) \cite{Hong-Choi} and those calculated from an elaborate dynamical mean-field calculation of the Hubbard model \cite{Gull-Millis2015, Gull-Millis2014}. They give neither the principal features of the normal nor of the pairing self-energy.

The experimental results may be summarized with the conclusion that at $T\approx T_c$,
\be
\label{result1}
I_{irr}^{ph}({\bf k,k}', \omega-\omega') \approx - \frac{I_{irr}^{pp}({\bf k,k}', \omega-\omega')}{\cos(2\theta_{\bf k})\cos(2\theta_{\bf k}')} \approx g_0^2 N(E_F) C(\omega-\omega').
\ee
$C(\omega-\omega')$ is consistent with the quantum-critical spectra of the 2D-DQXY model, as is its separable form in momentum and frequency. It is also consistent with the vertex of the form (\ref{coupling3}). Eq. (\ref{result1}) ignores the bump at around 50 meV in the spectra in $I_{irr}^{ph}$, which is absent in $I_{irr}^{pp}$, and which
from measurements of relaxation rates by pump-probe optical experiments \cite{vanderMarel2012} is deduced to be of different origin than the quantum-critical spectra. (\ref{result1}) also ignores the observed angular anisotropy of the normal single-particle self-energy, discussed above.

One may construct the complete vertex from the irreducible vertices using the Bethe-Salpeter equations \cite{Nozieres-book, AGD}. In the approximation that the single-particle self-energies are momentum independent, it is easy to see that the singularities of the complete vertex are the same as of the irreducible vertices. The weak angular dependence makes the calculation harder but cannot change the singularities. Knowledge of the complete vertex solves the problem.

\subsubsection{Superconductivity in the Fe-based compounds}

Considering the region of its occurrence, superconductivity in the Fe-based compounds (and in the heavy fermion compounds) is undoubtedly promoted by AFM fluctuations. The predictions for the normal and pairing self-energies if the fluctuations (and the coupling functions) can be obtained from those of the 2D XY model, and are quite different from those from the traditional theory of promotion of superconductivity by AFM fluctuations \cite{MiyakeSV1986, Scalapino1986}. For the normal self-energy, the prediction in the quantum-critical region is that it is linear in $max(\omega, T)$ and momentum independent just as in the cuprates. This would also explain the linear in T resistivity, and the other anomalies. But it ought to be borne in mind that there are often significant 3D couplings in these materials as well as higher energy cut-offs due to additional physics in many of the Fe-based compounds. So, the regime of occurrence of quantum-criticality may not be as clear and wide as in the hole-doped cuprates.

 The observation of pairing in some of these compounds in which there is no nesting of electron and hole Fermi-surfaces appears to remove for them (and by implication, for others) the weak-coupling mechanism for either antiferromagnetism or for pairing due to the traditional form of fluctuations, as has been noted \cite{Dung-Hai2009, Ab-Si2011}.

These compounds however also have very unusual parameters \cite{Hosono-rev, Kotliar-Hund, Dung-Hai2009}, besides having many bands crossing the Fermi-surface. For example, the bottom of the conduction band measured from the Fermi-energy is often less than 0.1 eV, which is similar to  the upper cut-off of the antiferromagnetic fluctuations \cite{Dai-RMP} and much smaller than the interaction energies. It is possible that they may be paired in {\it amplitude} already in the normal state \cite{Matsuda_Localpairs}. This issue is also connected with the remarkable fact that the uniform magnetic susceptibility of these compounds decreases as temperature decreases \cite{Hosono-rev} and that the relation of the specific heat at the transition to the background specific heat \cite{Canfield_Fe_spht} is quite unlike that given by BCS class of  theories. These are among the prominent new questions posed by these compounds which await further investigations.

The self-energy of the Fe-based superconductors in the normal and the superconducting state have not yet been deduced by experiments. We expect that for 2D class of such compounds, the normal self-energy in the quantum-critical fluctuation regime, is again $\propto max(\omega, T)$ at all angles around any given fermi-surface.  The symmetry of the superconducting state appears to vary depending on the compound and reflects probably the complications due to multi-band nature or to features not yet understood, due possibly to the unusual parameters (Fermi-energy smaller or at the same scale as $\omega_c$) in these compounds.\\

{\it Acknowledgements:} The work reviewed here was partially supported by the National Science Foundation under grant DMR DMR 1206298. This review was written while on a sabbatical at MIT. I am thankful to Patrick Lee for organizing it. I wish to thank Vivek Aji, Han-Yong Choi, Xingjiang Zhou and Lijun Zhu for collaborations which led to this review and for useful discussions. Discussions with various experimental groups around the world, too numerous to name were essential to the work surveyed here.

\bibliographystyle{naturemag}
\bibliography{ms2.bbl}

\end{document}